\documentclass[epj]{svjour}
\usepackage{graphicx}
\usepackage{amssymb,amstext}
\usepackage{url}
\def\aap{Astron. Astrophys.}%
\def\apj{Astrophys.\ J.}%
\def\apjs{Astrophys.\ J.\ Suppl.}%
\def\apjl{Astrophys.\ J.}%
\def\prc{Phys.\ Rev.\ C}%

\begin{document}
\title{Have Superheavy Elements been Produced in Nature?}
\author{I. Petermann\inst{1}
\and K. Langanke\inst{2,3,4}
\and G. Mart\'inez-Pinedo\inst{3,2}
\and I. V. Panov\inst{5,6}
\and P.-G. Reinhard\inst{7}
\and F.-K. Thielemann\inst{5}}
\offprints{}          
\institute{Argelander-Institut f\"ur Astronomie, Universit\"at Bonn, Auf dem
H\"ugel 71, D-53121 Bonn, Germany \and GSI Helmholtzzentrum f\"ur Schwerionenforschung,
  Planckstra{\ss}e 1, D-64291 Darmstadt, Germany 
\and Technische Universit\"at Darmstadt, Institut f\"ur
  Kernphysik, Schlossgartenstr. 2, D-64289 Darmstadt, Germany 
\and Frankfurt Institute for Advanced Studies, Frankfurt,
  Ruth-Moufang Str. 1, D-60438 Frankfurt, Germany \and Department of
  Physics, University of Basel, Klingelbergstr. 82, 4056 Basel,
  Switzerland \and Institute for Theoretical and Experimental Physics,
  B. Cheremushkinskaya St. 25, 117259 Moscow, Russia \and Institut
  f\"ur Theoretische Physik II, Universit\"at Erlangen-N\"urnberg,
  Staudtstrasse 7, D-91058 Erlangen, Germany}  
\date{\today}
%
\abstract{
We discuss the possibility whether superheavy elements can be produced
in Nature by the astrophysical rapid neutron capture process. To this
end we have performed fully dynamical network r-process calculations
assuming an environment with neutron-to-seed ratio large enough to
produce superheavy nuclei. Our calculations include two sets of
nuclear masses and fission barriers and include all possible fission
channels and the associated fission yield distributions. Our
calculations produce superheavy nuclei with $A\approx 300$ that
however decay on timescales of days. 
} 
\maketitle
\section{Introduction}
\label{intro}

More than 40 years ago, the possible existence of an island of
superheavy nuclei was proposed (see e.g.
Refs.~\cite{MyersSwiatecki:1966,Sobiczewski:etal:1966,Meldner:1969}),
due to the appearance of shell closures of protons and neutrons at
mass numbers beyond the then available nuclear data.  Various models
predicted shell closures at $Z=114$ and $N=184$
\cite{Meldner:1969,Nilsson:etal:1969a,Nilsson:etal:1969b,Grummann:etal:1969,Brueckner:etal:1971}.
More recent microscopic calculations
\cite{Cwiok:etal:1996,Bender:etal:1999,Kruppa:etal:2000,HeenenNazarewicz:2002,Cwiok:etal:2005}
suggest that the magic proton number could be higher than $Z=114$
(possibly Z=120, 124 or 126) and the neutron shell closure occurs at
$N=172$ or 184. In fact, shell closures are somewhat washed out to
regions of shell stabilization around these
numbers~\cite{Bender.ea:2001}. Shell closures make superheavy nuclei
stable against the dominating decay channel in this region:
spontaneous fission.  The highest fission barriers are expected for
nuclei just above the closed proton shell ($Z=114$) and just below the
closed neutron shell at $N=184$. This is of high relevance for the
decay half-life of beta-stable nuclei in this mass
region~\cite{MollerNix:1994,Moller:etal:2009,Moller:2010,SobRoz:2011,Erler.Langanke.ea:2012}.

Over the last thirty years, several experimental campaigns at
Berkeley, GSI, RIKEN, and JINR Dubna have explored this predicted
island of superheavy elements of increasingly stable nuclei around
atomic number 114 (see
e.g. \cite{HofmannMuenzenberg:2000,Armbruster:2000,Oganessian:2004,Moody:etal:2004,Hofmann:2009})
and have found by now elements with charge numbers up to $Z=118$
\cite{Oganessian.etal:2006,Oganessian:etal:2010,Oganessian:etal:2011}.
The decay chains observed so far suggest the existence of a region of
enhanced stability created by shell effects. Alpha-decay chains with
fission, occurring at the end of the chain, explored the boundary of
the island at low Z and N. The question remains now how far the island
extends in the other directions (low Z, high N; high Z, low N; high Z,
high N) and what experimental paths can lead us to these regions.

A related question has been addressed almost as long, i.e. whether
superheavy elements are/could be produced in
Nature~\cite{SchrammFowler:1971,Flerov.Ter-Akopian:1983}.  The only
candidate process for astrophysical nucleosynthesis of these heavy
elements is the r(apid neutron capture) process (for reviews see
\cite{Cowan:etal:1991,Arnould:etal:2007,Thielemann:etal:2011}).
Alternative terrestrial, non-accelerator, experiments have been
discussed~\cite{Meldner:1972,DornHoff:1965,Bell:1967,Igley:1969,Becker:2004}.

While the very early discussion made it plausible that superheavy
elements can be produced in r-process
nucleosynthesis~\cite{SchrammFowler:1971}, more systematic
approaches, including neutron-induced fission estimates, predicted
that the r-process path would reach areas dominated by fission and the
path to superheavy elements would be blocked in nature, dependent,
however, on mass model
uncertainties~\cite{Boleu:etal:1972,SchrammFizet:1973,Brueckner:etal:1973,HowardNix:1974}.

The first extended table of fission barrier predictions by Howard and
M\"oller \cite{HowardMoller:1980} permitted to engage in more
systematic calculations.  Thielemann, Metzinger and Klapdor
\cite{Thielemann:etal:1983a} performed beta-delayed fission
calculations for a number of mass models in combination with the above
fission barrier predictions. In their application to r-process
calculations \cite{Thielemann:etal:1983b}, they came to the conclusion
that the r-process path ends in a 100\% beta-delayed fission region
and superheavies cannot be reached. Their results were based on
a combination of the Hilf mass formula \cite{Hilf:etal:1976} (which
contains a slightly too steep mass parabola) and the Howard and
M\"oller~\cite{HowardMoller:1980} fission barriers, which by now turn
out to be somewhat underestimated~\cite{Mamdouh:etal:2001}. 
The combination of both effects led
to an overestimation of the effect of
fission~\cite{Cowan:etal:1991} in the r-process path around
$Z=92$. 

\begin{figure}
  \includegraphics*[width=\linewidth]{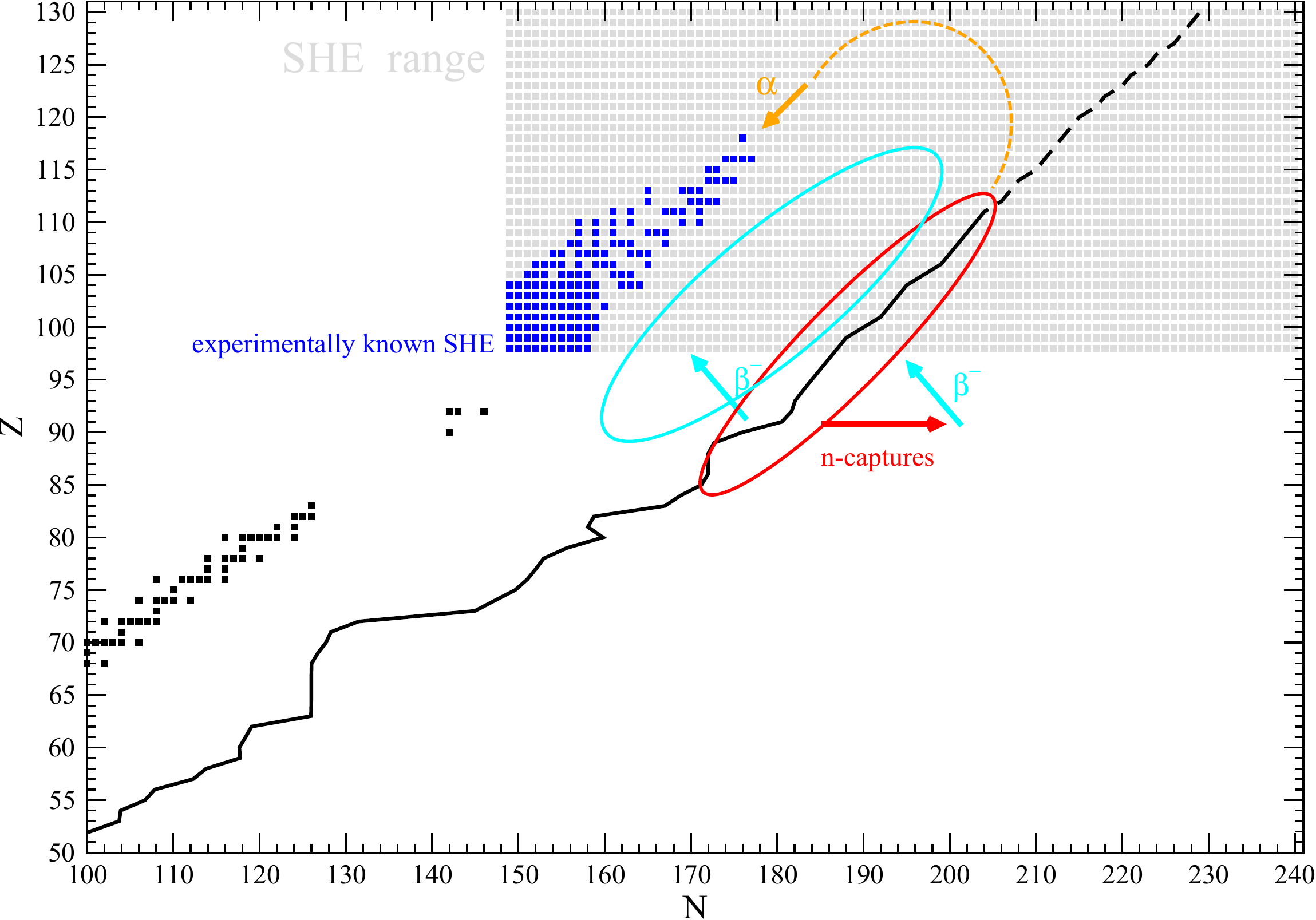}
  \caption{Possible pathways of the r-process in the nuclear chart
    (solid line). On the path  and the
    subsequent  decay back to beta-stability there exist three options to
    encounter or avoid fission on the way to superheavy nuclei (options i)-iii)     in the text).
    in option iii) superheavies  would be made by producing
    progenitors with even higher charge numbers which then decay
    by subsequent beta- and alpha-decays. The
    present investigation, limited to network calculation up to Z=110,
    cannot explore this latter option. \label{fig:i-iii}}
\end{figure}

Since then, after considering these deficiencies, studies aimed at
using r-process nuclei for cosmochronometry have neglected
fission~\cite{Cowan:etal:1999} and did not address the question of
production of superheavy nuclei.  In recent years improved predictions
of fission barrier tabulations for extended ranges of nuclei have
become
available~\cite{Mamdouh:etal:1998,MyersSwiatecki:1999,Moller:etal:2009,Goriely:etal:2009},
being based on the Extended Thomas-Fermi (ETFSI), Thomas-Fermi (TF),
Finite Range Droplet Model (FRDM) or the Hartree-Fock Bogoliubov (HFB)
approach. This also led to new neutron-induced fission and
beta-delayed fission
calculations~\cite{Panov:etal:2005,Panov:etal:2010} that permitted to
address the formation of superheavy elements and long-lived
cosmochronometers. In order to explore these issues we have performed a new
set of r-process calculations with few preliminary results reported
in~\cite{Panov:etal:2009,Petermann:etal:2010,Langanke:etal:2011}. In
this manuscript, we present detailed calculations aiming to answer the
following questions: (i) is the r-process experiencing strong fission
effects during the build up of heavy nuclei by neutron captures and
beta-decays or can unstable extremely neutron-rich nuclei with mass
numbers up to $A=300$ and beyond be produced? (ii) Does the subsequent
decay to beta-stability pass by areas of the nuclear chart where
fission dominates, hence blocking the production of long-lived
superheavy nuclei? (iii) Can this area of fission dominance be
surpassed during the r-process towards higher charge numbers, so that
subsequent beta-decay ends with finite abundances in the valley of
beta-stability, followed by alpha-decay chains towards the island of
superheavy nuclei? 
This three options/questions are indicated in Fig.~\ref{fig:i-iii}.

Our paper is structured in the following way. Section 2 discusses the
nuclear input used: masses and fission barriers of nuclei far from
stability. Section 3 discusses the r-process model and presents the
results of the nucleosynthesis calculations. Section 4 analyzes 
the results addressing the reliability of the nuclear input due
to comparison with data. Section 5 presents the conclusions.

\section{The Role of Masses and Fission Barriers}
\label{sec:1}

If one wants to address the question whether the island of
superheavy elements was reached in natural neutron capture
environments, like final products of r-process nucleosynthesis, a
complete knowledge of all possible reaction and decay properties is
required. This includes neutron capture, photo-disintegration,
neutron-induced fission, beta-decay, beta-delayed fission, spontaneous
fission
and for sufficiently large temperatures gamma-induced fission.  In
order to provide this information in unknown territory, theoretical
prediction are required. Prediction of these properties rely therefore
on masses (to determine reaction Q-values), fission barriers, optical
potentials for transmission coefficients of particle channels,
transmission coefficients for gamma-tran\-sitions (determined either
from giant resonance properties or microscopic approaches), level
densities of excited states (obtained from back-shifted Fermi gas or
combinatorial
approaches)~\cite{Goriely:etal:2009,Panov:etal:2010,RauscherThielemann:2000,Mocelj:etal:2007,Loens:etal:2008,Litvinova:etal:2009,Cyburt:etal:2010,Rauscher:2011,Goriely:etal:2008,LarsenGoriely:2010,Goriely:etal:2011}.
Spontaneous fission has been described with different levels of
sophistication including: a simple formula fitted to experimental data~\cite{KodamaTakahashi:1975,RenXu:2005}, 
macroscopic-microscopic
approaches~\cite{Moller:etal:2009,Smolanczuk:1995,Smolanczuk:1997}, and
Skyrme-Hartree-Fock approaches~\cite{Erler.Langanke.ea:2012,Goriely:etal:2009}.

For a number of years extended and increasingly reliable mass
predictions have been available, the Finite Range Droplet Model (FRDM)
\cite{Moller:etal:1995}, the Extended Thomas Fermi Model with
Strutinski Integral (ETFSI) \cite{Aboussir:etal:1995}, the
``microscopic'' Duflo-Zuker mass formula \cite{DufloZuker:1995}, and the
Har\-tree-Fock-Bogol\-iubov approach with Skyrme
functionals~\cite{Goriely:etal:2009b,Goriely:etal:2010}. Relativistic
and non-relativistic mean field approaches 
which do not yet reach an accuracy level in ground state masses of
0.6--0.7 MeV have not been considered in r-process simulations.

\begin{figure}
  \includegraphics[width=\linewidth]{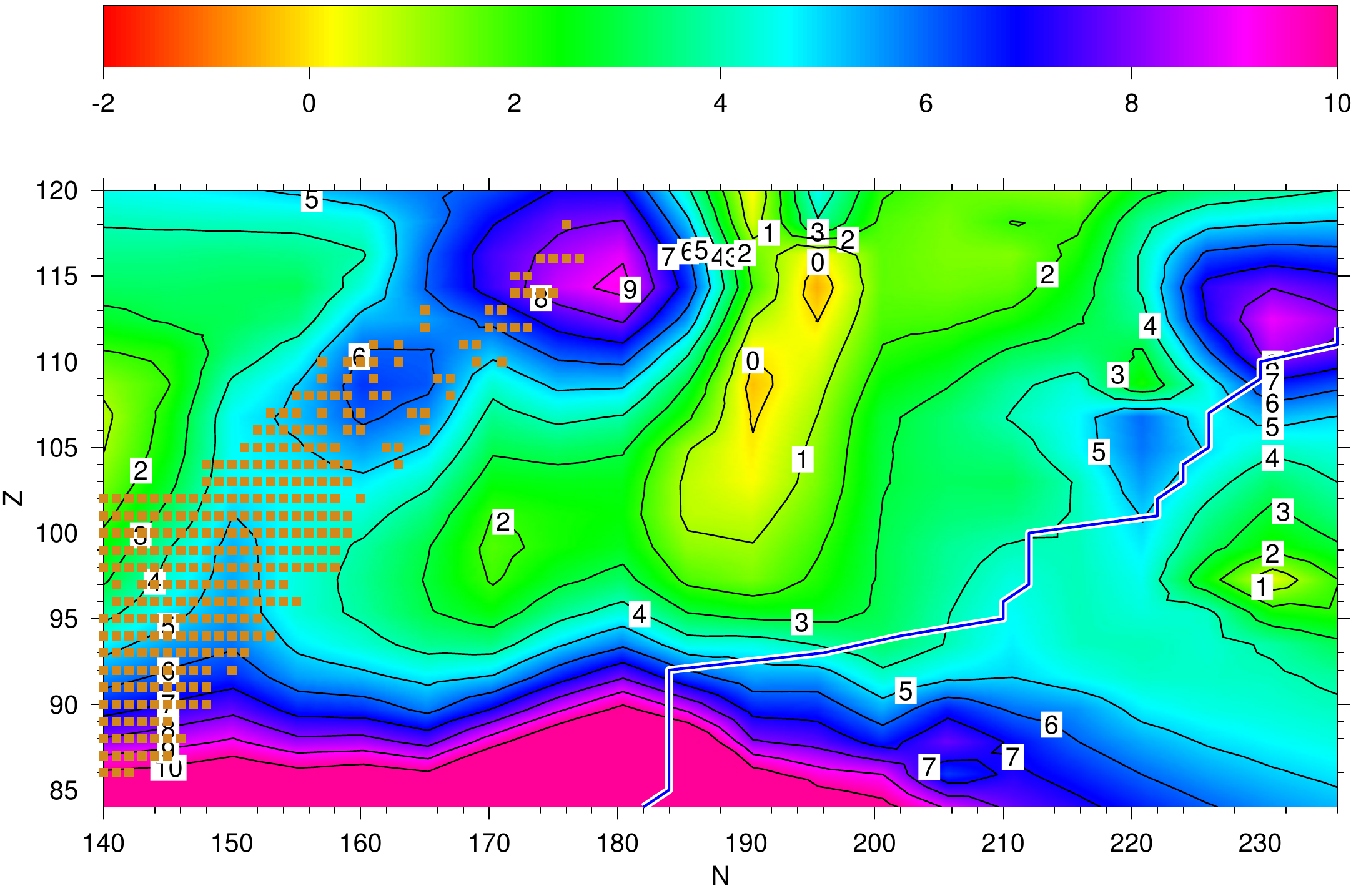}\\
  \includegraphics[width=\linewidth]{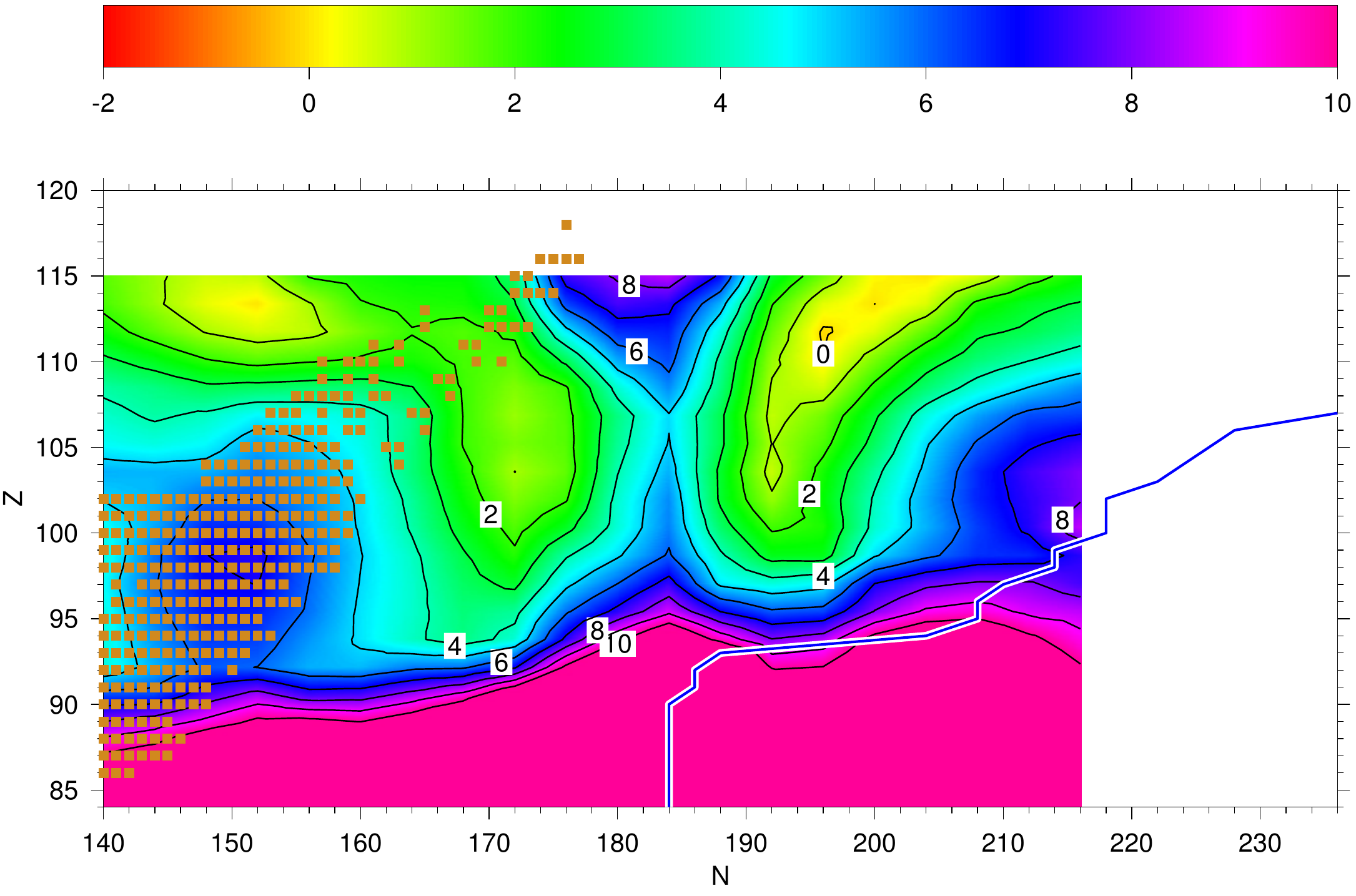}
  \caption{Contour plots of the largest fission barrier heights from
    the TF (top)~\cite{MyersSwiatecki:1999} and ETFSI
    (bottom)~\cite{Mamdouh:etal:1998} mass models. This figure also
    indicates to which extent data sets are available in the nuclear
    chart. HFB barriers are only publicaly available in the rage
    $Z=90$--102~\cite{Goriely:etal:2009} and FRDM
    barriers~\cite{Moller:etal:2009} only close to
    stability.\label{fig:barriers}} 
\end{figure}

Global predictions of fission barriers have been scarce.  After the
Howard and Moller~\cite{HowardMoller:1980} predictions (which are now
known to produce too small barriers), new attempts have only been
performed in the last
decade~\cite{Moller:etal:2009,Erler.Langanke.ea:2012,Mamdouh:etal:1998,MyersSwiatecki:1999,Goriely:etal:2009,Goriely:etal:2011,Sheikh.Nazarewicz.Pei:2009}. 

They are based on the ETFSI model~\cite{Mamdouh:etal:1998}, the
Thomas-Fermi approach~\cite{MyersSwiatecki:1999}, with a similar
macroscopic mass para\-bola as FRDM, combined with shell corrections of
the Finite Range Liquid Drop model FRLDM~\cite{Moller:etal:2009} -
similar, but not identical to FRDM, and the
Skyrme-Hartree-Fock-Bogoliubov
model~\cite{Goriely:etal:2009,Goriely:etal:2011} based on the density
functional used for the HFB-14
masses~\cite{Goriely.Samyn.Pearson:2007}. The Skyrme Hartree-Fock
approach has also been used for a sensitivity study of fission
barriers for superheavy nuclei~\cite{Erler.Langanke.ea:2012} and for
the dependence of the fission barrier on excitation
energy~\cite{Sheikh.Nazarewicz.Pei:2009}. 

Fig.~\ref{fig:barriers} shows contour plots of the largest fission
barrier for individual nuclei from the TF (top) and ETFSI (bottom)
models. The calculation of beta-delayed or neutron-induced fission
requires the knowledge of mass differences that should be derived
consistently from the same model as the
barriers~\cite{Cowan:etal:1999}. Unfortunately this is not always
possible, however based on the similarity of the underlying model we
expect that the most reasonable combinations for barriers and masses
are ETFSI/ETFSI, TF/FRDM, FRLDM/FRDM, and HBF-14/HFB-14.
Of these four sets only two (ETFSI/ETFSI, TF/FRDM) can currently be
used for global r-process calculations which require knowledge of
nuclear properties from stability to the neutron drip line for an
extended range of Z values. The Bruslib database of astrophysical
reaction rates~\footnote{\url{http://www.astro.ulb.ac.be/bruslib}}
includes fission barriers based on the HFB-14 model for the range
$Z=90$--110 and neutron induced rates up to $Z=102$. The recent FRLDM
barriers~\cite{Moller:etal:2009} cover the range $Z=78$--125 but
always close to the beta-stability line. 

\begin{figure}
  \includegraphics[width=\linewidth]{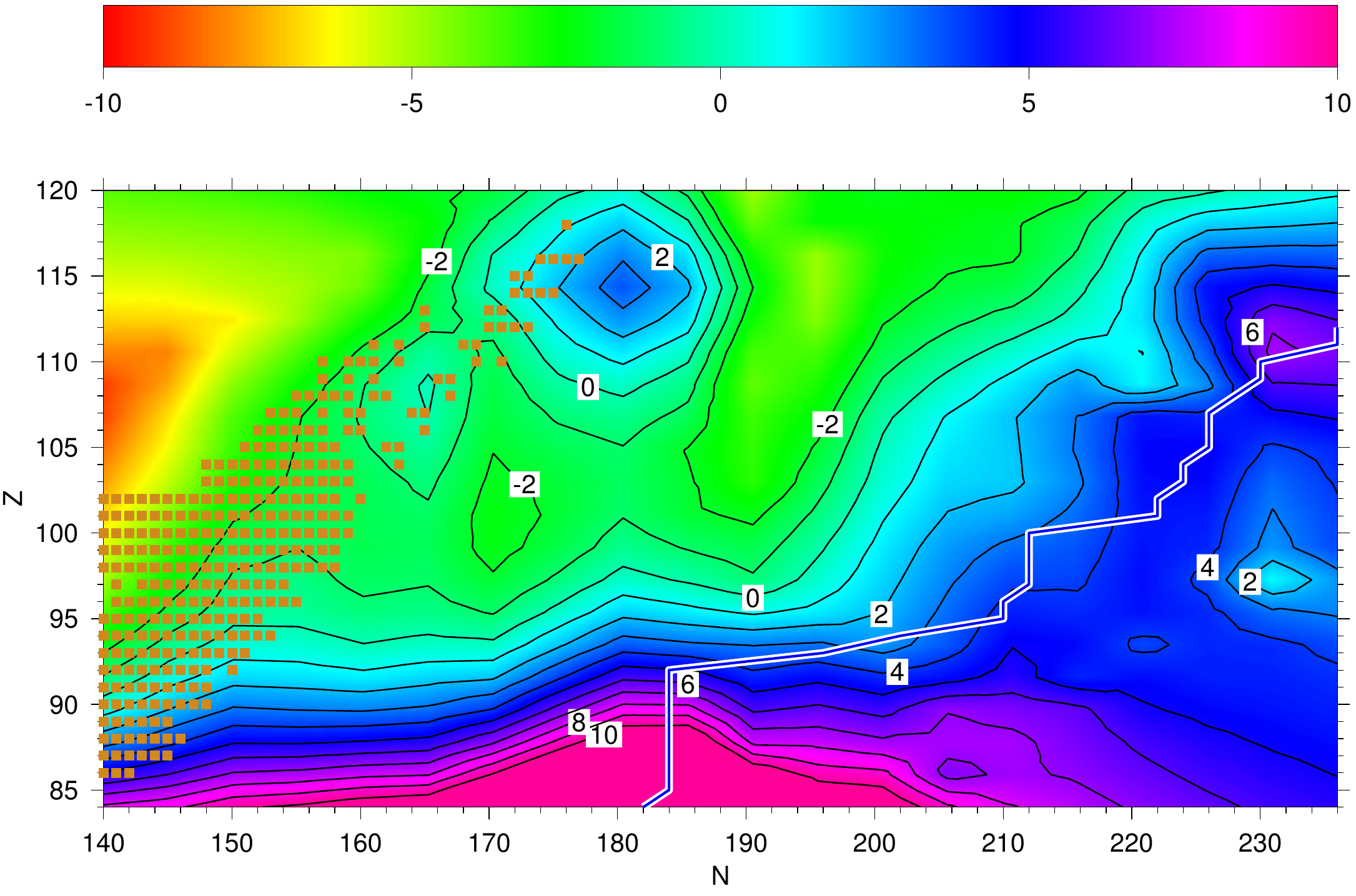}
  \includegraphics[width=\linewidth]{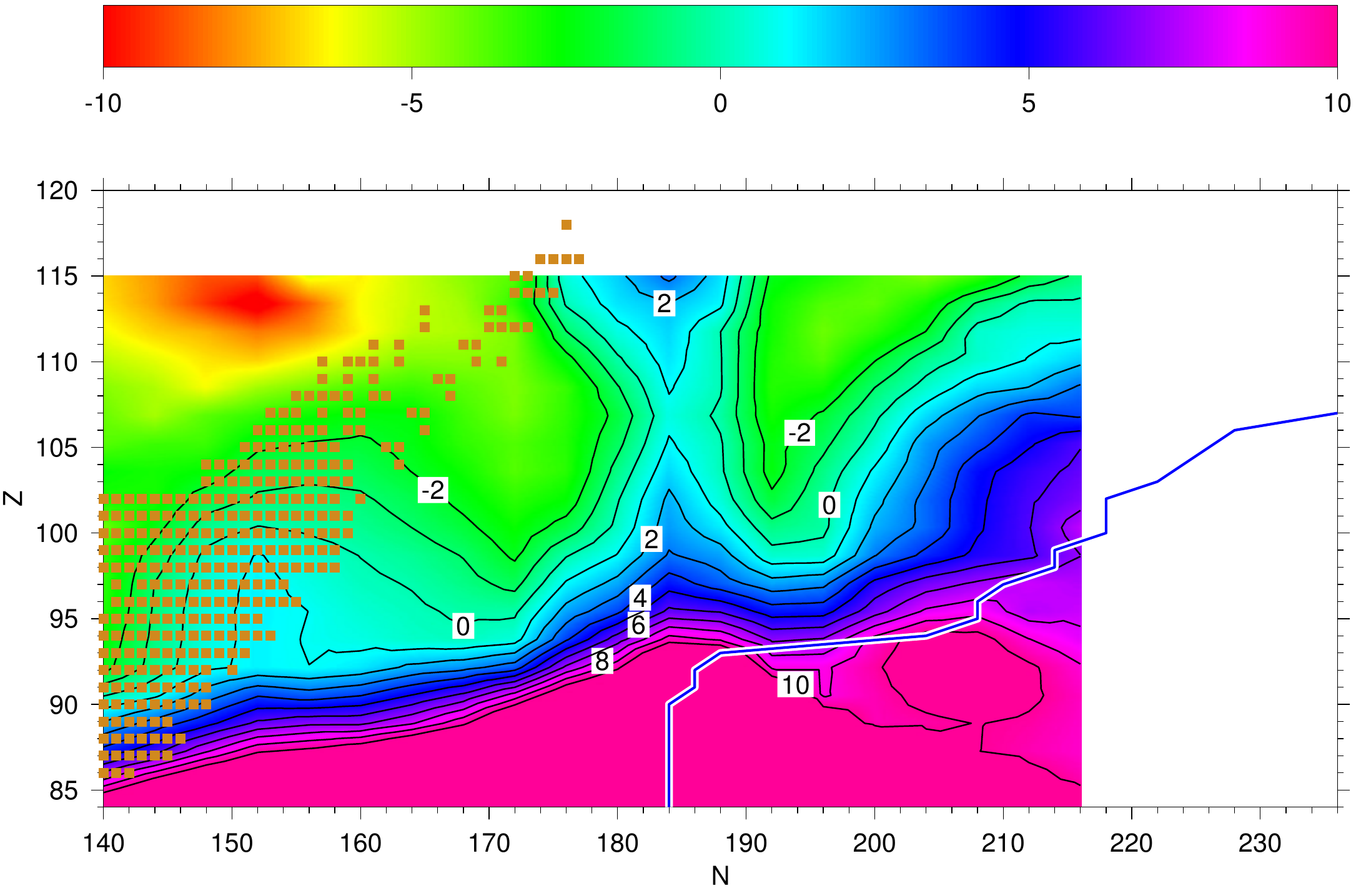}
  \caption{Contour plots of $B_f-S_n$ for the combinations TF/FRDM
    (top) and ETFSI/ETFSI (bottom). While there exist positive values
    of $B_f-S_n$ for neutron numbers around $N=184$ also for $Z>95$ in
    the bottom figure (indicating that neutron-induced fission is not
    permitted), this is not the case for TF/FRDM. The neutron-drip
    line is shown for the appropriate mass modes (FRDM and ETFSI). It
    can be seen that in the vicinity of the neutron-drip line
    neutron-induced fission is clearly inhibited.\label{fig:bfsn}}
\end{figure}

\begin{figure}
  \includegraphics[width=\linewidth]{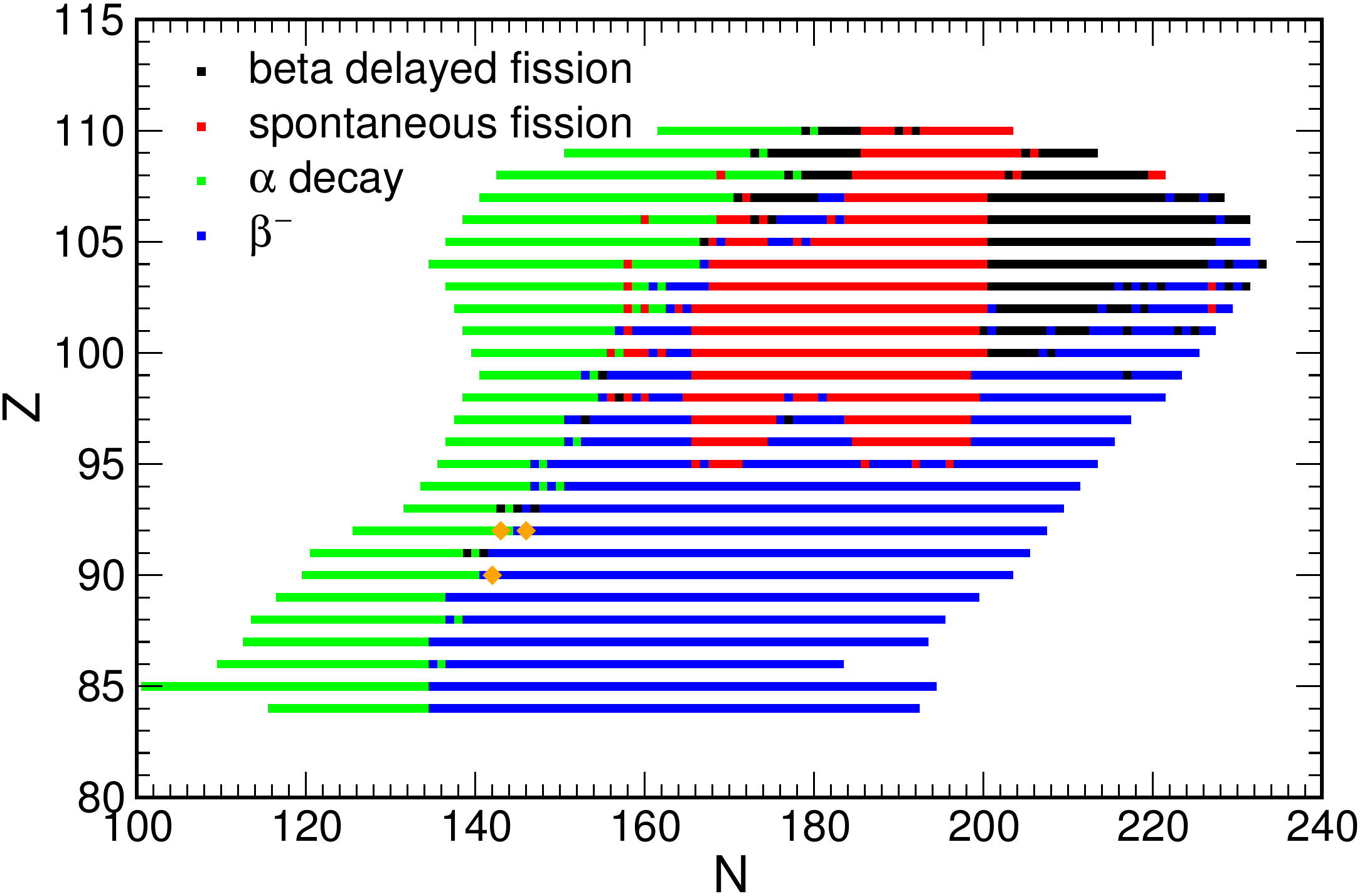}\\
  \includegraphics[width=\linewidth]{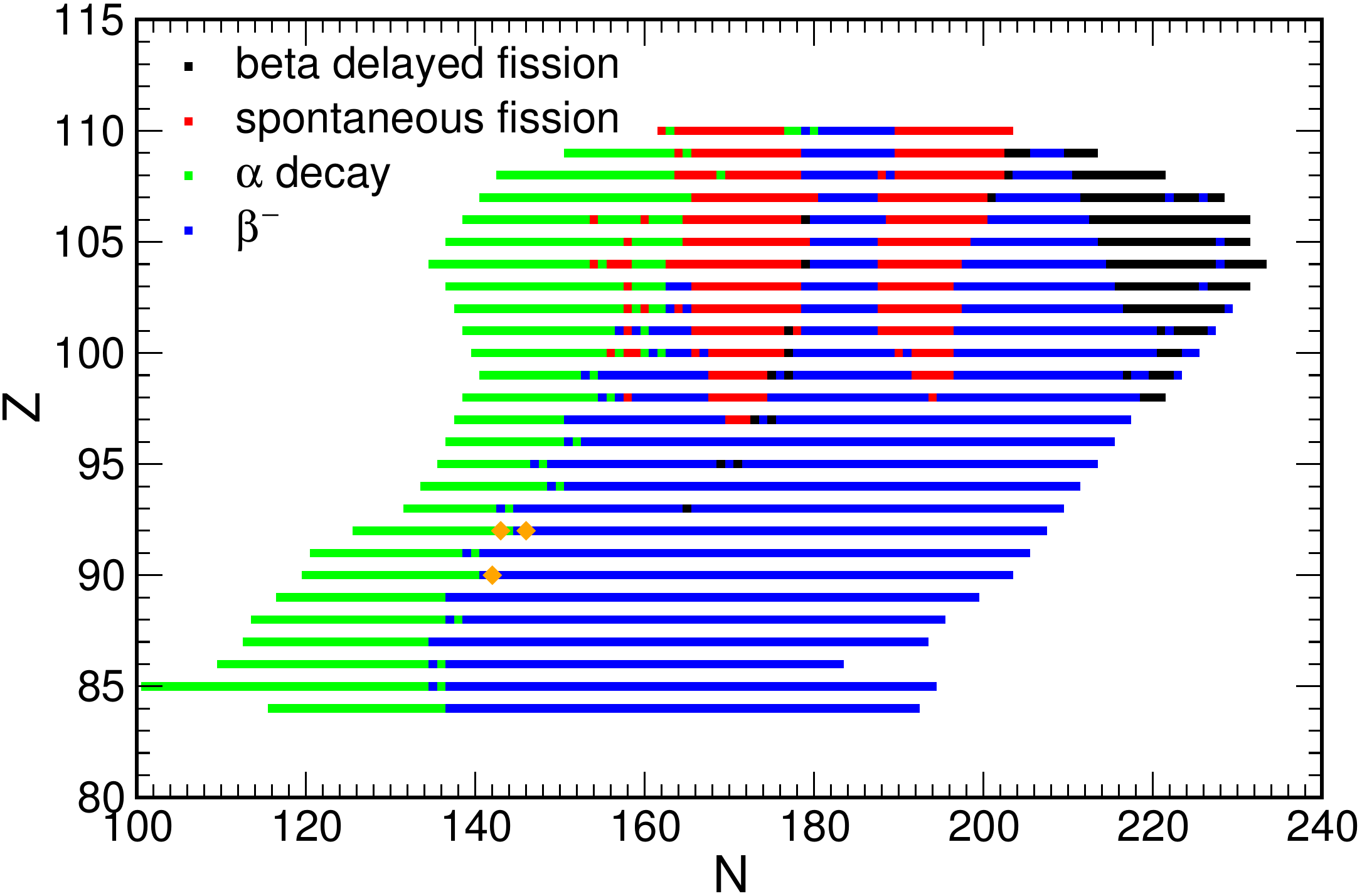}
  \caption{Dominant decay channels: alpha-decay on the proton-rich
    side of stability - not of importance here, beta-decay,
    beta-delayed fission in regions with positive
    $Q_\beta-B_f(daughter)$ and regions where spontaneous fission
    half-lives are shorter than beta-decay half-lives.\label{fig:dominant}} 
\end{figure}

In figures~\ref{fig:bfsn} and~\ref{fig:dominant} we present the
different reaction and decay properties for neutron-induced,
beta-delayed and spontaneous fission. Figure~\ref{fig:bfsn} shows
contour plots of the difference between fission barrier and neutron
separation energy, $B_f - S_n$. Nuclei for which this value is
below~2~MeV are expected to fission immediately after capturing a
neutron. Given the fact that fission barriers do not suffer from
strong odd-even effects while the neutron separation energies are
affected by those, we expect that neutron-induced fission
will mainly occur for even-N r-process nuclei.


Figure~\ref{fig:dominant} shows the dominant decay channel
(alpha-decay on the neutron-deficient side of stability, beta-decay,
beta-delayed fission in regions where the beta-decay $Q$-value is
larger than the fission barrier of the daughter nucleus, and
spontaneous fission), see
also~\cite{Panov:etal:2005,Panov:etal:2010,Langanke:etal:2011,Martinez-Pinedo:etal:2007}.
As the TF barriers provide only information about the height of the
largest barrier we have estimated the spontaneous fission half-lives
using the simple parametrization of
Ref.~\cite{KodamaTakahashi:1975}, but with coefficients adjusted to
the available experimental fission half-lives related to the TF
fission barriers:

\begin{equation}
  \label{eq:sphalf}
  \log t_{1/2} (\text{s}) = 8.08 B_f - 24.05,
\end{equation}
where the barrier $B_f$ is given in MeV.  We use the same parameters
for the ETFSI barriers so that for the same barrier both models
predict the same spontaneous fission half-life.  Alpha-decay
half-lives are obtained using the Viola-Seaborg semi-empirical
relationship between half-life and alpha decay energy with parameters
determined in ref.~\cite{Sobiczewski.Patyk.Cwiok:1989}. We use
beta-decay half-lives and beta-delayed neutron emission probabilities
from ref.~\cite{Moeller.Pfeiffer.Kratz:2003} which are currently the
only values available for the mass range of interest here. As these data
only extend to $Z=110$, this is the upper limit for our r-process
network. Whenever an experimental half-life is known we use this value
in our calculations.

Beta-delayed fission rates are determined based on the FRDM beta-decay
rates~\cite{Moeller.Pfeiffer.Kratz:2003} using an approximate strength
distribution for each decay based on the neutron-emission
probabilities. For each fissioning nucleus and fission channel the
fission yields, including the amount of neutrons produced, are
determined~\cite{Zinner:2007} using the statistical code
ABLA~\cite{Gaimard.Schmidt:1991,Benlliure.Grewe.ea:1998}.

While there are obvious differences for fission barrier predictions of
the different models shown in Fig.~\ref{fig:barriers}, they become
more prominent in $B_f - S_n$ (Fig.~\ref{fig:bfsn}) and the prominent
decay channels as shown in Fig~{\ref{fig:dominant}}, especially close
to the neutron shell closure $N=184$. The content of
Figs.~\ref{fig:bfsn} and~\ref{fig:dominant} provides the information
for judging which decay/reaction channel is dominant in a particular
region. However, the determination of the neutron-induced fission rate
requires also the knowledge of the neutron density,
$n_n$, of an r-process environment. One important result is that
$B_f-S_n$ is always positive close to the neutron-drip line,
independently of the mass model. This is expected as the fission
barriers increase while the neutron separation energies decreases towards
the neutron drip line. This means that an r-process does not
experience strong neutron-induced fission if proceeding close to the
drip line for $N<184$. However, at the $N=184$ magic shell the drip
line moves closer to beta-stability and the quantity $B_f-S_n$ becomes
smaller.
For the TF/FRDM model, its value is around 2~MeV for nuclei around
$Z\sim 94$, $N\sim 186$ for which the neutron-induced fission
dominates over the $(n,\gamma)$ channel. This inhibits the production
of heavier nuclei, due to neutron-induced fission. The ETFSI/ETFSI
model predicts values of $B_f-S_n \approx 8$~MeV in the same region,
remaining large along the r-process path to heavier nuclei. The build
up of heavy nuclei continues and only ends by beta-delayed fission in
the region $Z\sim 105$, $N\sim 220$, see black region in
figure~\ref{fig:dominant}. Notice that both models predict very
similar neutron separation energies but rather different fission
barriers.

In the range of elements up to $Z=115$, regions dominated by
spontaneous fission will always be encountered during the decay to
beta-stability. For the TF/FRDM case this is unavoidable, as an
extended and connected region dominated by spontaneous fission has to
be passed. In the ETFSI/ETFSI case, the spontaneous fission region is
split into two ``islands'' divided by a range of about 10 units in
neutron number around $N=184$, where beta-decay dominates without being
followed by delayed fission. Thus,
beta-decay back to stability will have the chance to proceed through
the channel between the islands, until it encounters the region with
$N<184$. Dependent on the beta-decay half-lives encountered,
superheavy nuclei could exist for an extended period of time before
fissioning. We cannot make statements for a possible production of
nuclei beyond $Z=110$ as this is the upper limit of our present
network calculations due to the restricted availability of the nuclear
input. For this region extended sets for theoretical predictions have
been published~\cite{Erler.Langanke.ea:2012} after finishing our
calculations and in the future we plan to explore their impact in
r-process nucleosynthesis.

The decay and reaction inputs discussed in this section and displayed
in Figs.~{\ref{fig:barriers}},~\ref{fig:bfsn} and \ref{fig:dominant}
provide the basis for applications to the r-process calculations given
in the following section.

\section{r-Process Calculations with fission}
\label{sec:2}

\subsection{Astrophysical Sites}

A high availability of neutrons, leading to a large ratio of neutrons
to seed nuclei (typically Fe-group nuclei or beyond, up to $A=80$) can
occur in two types of explosive astrophysical environments: (a) Very
neutron-rich material, which is obtained under high nuclear
matter/neutron star densities, where electron capture on protons by
degenerate electrons with high Fermi energies cause a proton-to-nucleon
ratio $Y_e$ of about 0.1, is ejected and leads to an r-process during
the expansion phase. Possible examples are the ejecta from neutron
star collision events,
\cite{LattimerSchramm:1976,Lattimer:etal:1977,Freiburghaus:etal:1999b,Goriely:etal:2011b},
or other environments ejecting highly neutronized material like fast
rotating core collapse supernovae with strong magnetic fields and jet
ejecta
\cite{Cameron:2003,Fujimoto:etal:2008,Winteler.Kaeppeli.ea:2012}. (b)
Matter which is only slightly neutron-rich but experiencing a fast
expansion and high entropies. In such a case the reactions
$^4\mathrm{He}(\alpha\alpha,n)^9$Be$(\alpha,n)^{12}$C and/or
$^4\mathrm{He}(\alpha\alpha,\gamma)^{12}$C are responsible to move
matter from $^4$He to $^{12}$C followed by a fast sequence of reactions
producing heavier seed nuclei and few neutrons remaining. At high
entropies, $\propto T^3/\rho$, three body reactions are suppressed due
to the low density and/or high number of photons hindering the
production of heavy seed nuclei and result in large amounts of
$^4$He (with $N=Z$) and a tiny amount of heavy seed nuclei. This
results in large remaining neutron-to-seed ratios. Examples of this case are
the neutrino-driven
wind~\cite{Woosley:etal:1994,Takahashi:etal:1994,QianWoosley:1996,Otsuki.Tagoshi.ea:2000,Thompson.Burrows.Meyer:2001,arcones.janka.scheck:2007,Farouqi.etal:2010,Arcones.Martinez-Pinedo:2011}
from the nascent proto-neutron star after supernova core collapse (provided that it can produce neutron-rich
ejecta~\cite{Fischer.etal:2010,Huedepohl.etal:2010}) and matter
ejected from accretion disks around black
holes~\cite{Surman:etal:2008,WanajoJanka:2011}.

In the above two options, the high neutron-to-seed ratio is the
dominant driving force behind a rapid neutron capture
(r-)process. However, dependent on the environment, the densities and
temperatures can vary and cause differences in the exact working of
the r-process. In order to produce nuclei as heavy as Uranium and Thorium which are only
produced by the r-process, one requires conditions with
neutron-to-seed ratios of the order of 150.

\subsection{r-Process model calculations}

The previous subsection outlined the still existing uncertainties in
astrophysical sites/conditions which will then also be affected
further by nuclear uncertainties discussed in
section~\ref{sec:1}. We stress that, independent of the actual
environment, the decisive feature of an r-process to produce superheavy
elements is a sufficiently high neutron-to-seed ratio. In the
following, we assume assumed conditions corresponding to 
fast expanding ejecta with high entropy, in order to ensure a large enough
neutron-to-seed ratio. 

We start our calculations at a temperature of 10~GK for which Nuclear
Statistical Equilibrium is applicable and assume that the matter
follows a homologous adiabatic expansion where the density 
behaves as:

\begin{equation}
  \label{eq:earlyt}
  \rho(t) = \rho_0 \exp(-t/\tau) . 
\end{equation}
We use $\tau = 3$~ms following the hydrodynamical simulations of
ref.~\cite{Arcones:etal:2007}. At later times hydrodynamical
simulations show that the evolution is not anymore homologous and can
be approximated
by~\cite{Meyer.Brown:1997,panov.janka:2009,wanajo.janka.kubono:2010}:

\begin{equation}
  \label{eq:latet}
  \rho(t) = \rho_1 \left(\frac{\Delta+t_1}{\Delta + t}\right)^2,
\end{equation}
where the parameter $\Delta$ represents the time scale on which the
matter evolves from conditions of almost constant density ($t<<\Delta$)
to constant velocity
($t>>\Delta$)~\cite{Arcones.Martinez-Pinedo:2011}. We use a value of
$\Delta = 2$~s in agreement with hydrodymanical
simulations~\cite{Arcones:etal:2007,Fischer.etal:2010}.

In order to attain a large neutron-to-seed ratio we assume a
constant entropy of 200~$k/\text{nucleon}$ and initial $Y_e = 0.35$
that results in a neutron-to-seed ratio of 290 at 3~GK. The
temperature is determined by the condition of constant entropy using
the equation of state of ref.~\cite{Timmes.Arnett:1999}.
Our network code is based on the XNet code of
ref.~\cite{Hix.Thielemann:1999} that has been extended to treat
implicitly all fission reactions. 

\begin{figure}[htb]
  \centering
  \includegraphics[width=\linewidth]{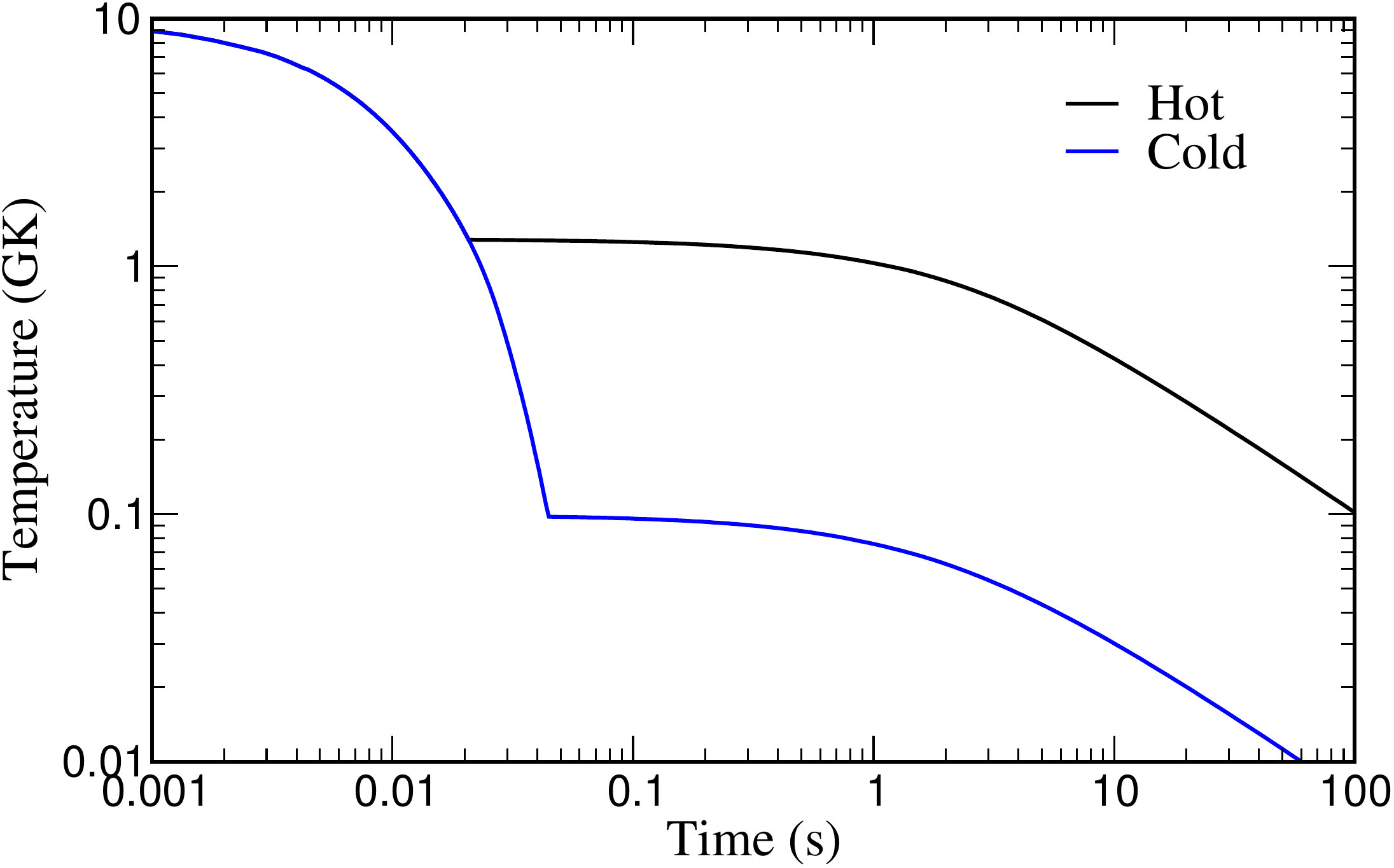}\\
  \includegraphics[width=\linewidth]{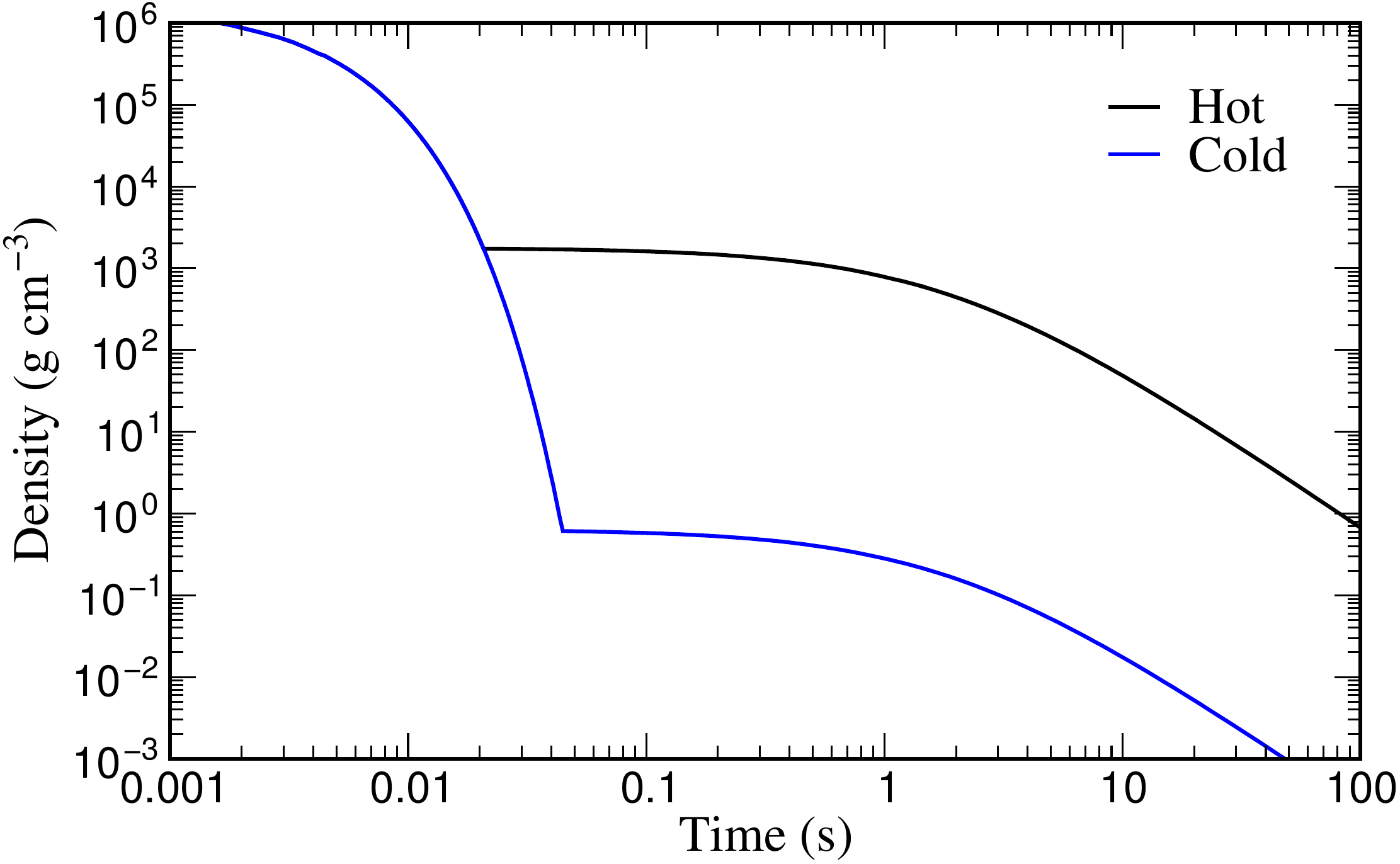}\\
  \caption{Evolution of temperature and density for a hot and cold r-process
depending on the choice of $t_1$ in Eq.(3). Notice that for both cases the
n/seed ratio drops below 1 in the range t=1-2s.}
  \label{fig:rhoT}
\end{figure}

As discussed in section 3.1, the r-process can occur under quite
different astrophysical conditions. In the following, we present two
extreme cases for the late time evolution (see figure~\ref{fig:rhoT})
by choosing $t_1$ as the time for which the temperature has reached
values of 1.15 and 0.1~GK, respectively. In the following, we refer to
these two cases as ``hot'' and ``cold'' r-process.

\begin{figure}[htb]
  \centering
  \includegraphics[width=\linewidth]{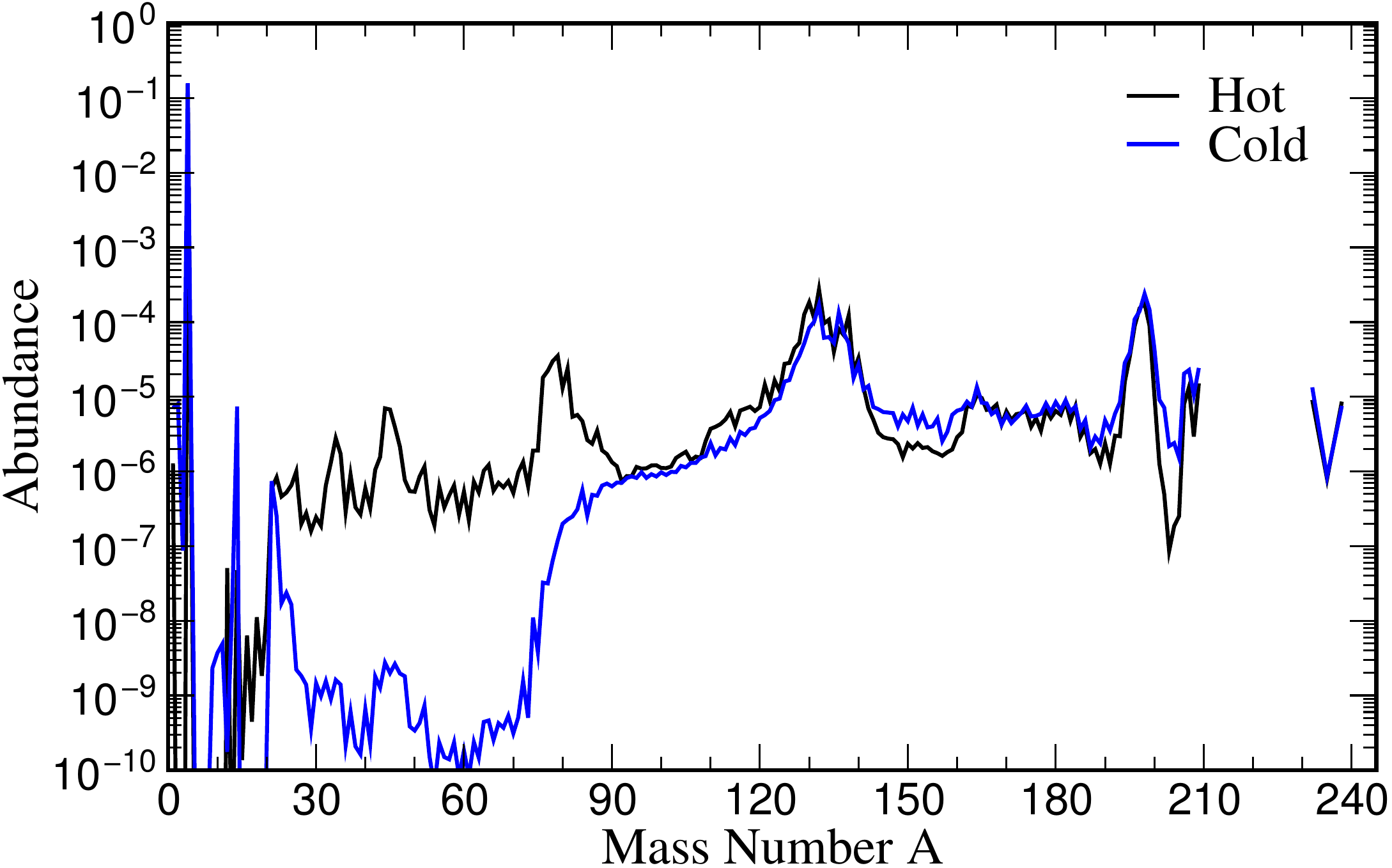}
  \caption{Final abundances for the hot and cold  r-process scenarios
    after 1 Gy; i.e. 
 after the decay of unstable isotopes 
 (see text for details).\label{fig:freezeout}}
\end{figure}

The nucleosynthesis in these two cases differs in two major aspects.
First, in the ``hot'' r-process calculations, we find that there is
a continuous production of seed nuclei during the whole duration of the
r-process as charged-particle reactions never completely freeze-out
(see ref~\cite{Sasaqui.Otsuki.ea:2006}). This is not the case in the ``cold''
r-process, as for such low temperatures charged-particle reactions are
too slow when compared with the dynamical timescale of r-process
nucleosynthesis. The continuous production of seed nuclei manifests
itself in enhanced abundances for nuclei in the range
$A=20$--90. This is demonstrated in figure~\ref{fig:freezeout} that
shows the final r-process abundances based on the TF/FRDM
input. Calculations with the ETFSI/ETFSI input show the same
features. 

The second difference relates to the role of photodisintegration
reactions during the r-process. At the larger temperatures of the hot
r-process the photodisintegration reactions are fast enough to
establish a $(n,\gamma)\rightleftarrows(\gamma,n)$ equilibrium. This
chemical equilibrium between both reaction types along an isotopic
chain, defines a narrow "r-process path", with a nucleus of
dominant abundance in each chain, acting as waiting point for
beta-decays. This is  the ``classical'' r-process picture as
introduced in ref.~\cite{Burbidge.Burbidge.ea:1957,Cameron:1957}. 

In the ``cold'' r-process, photodisintegration is too slow to
establish an equilibrium with neutron captures. The evolution is
rather determined by neutron captures and competing beta-decays which
opposite to photo-dissociation do not vary much from isotope to
isotope under r-process conditions. As a consequence several nuclei in
each isotopic chain are substantially populated which leads to a
broader r-process path. Calculations in the cold r-process scenario
have been already presented in~\cite{Blake.Schramm:1976,Wanajo:2007,Arcones.Martinez-Pinedo:2011},
however, with different aims than discussed here.

\begin{figure*}
  \centering
  \includegraphics[width=0.8\linewidth]{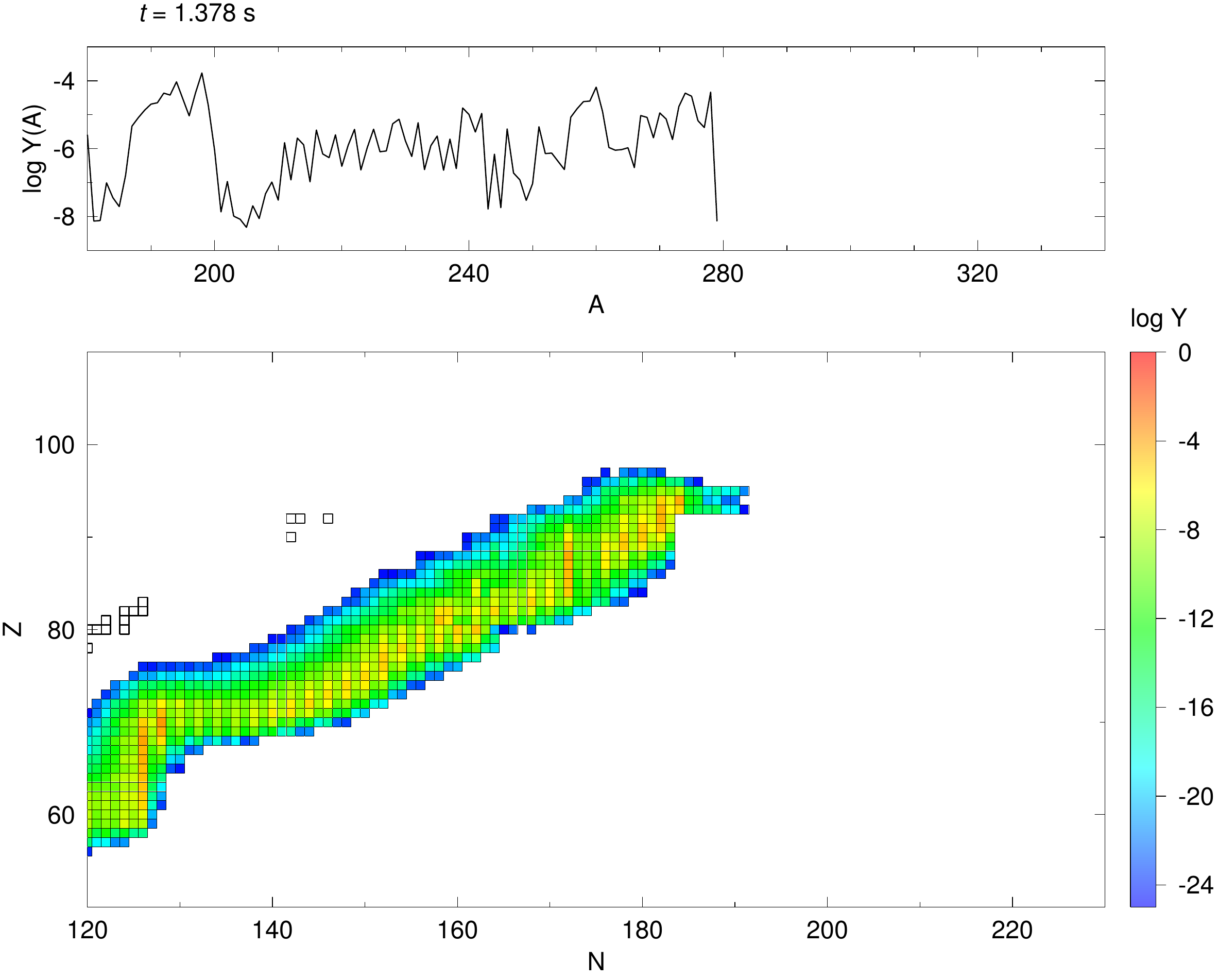}\\
  \includegraphics[width=0.8\linewidth]{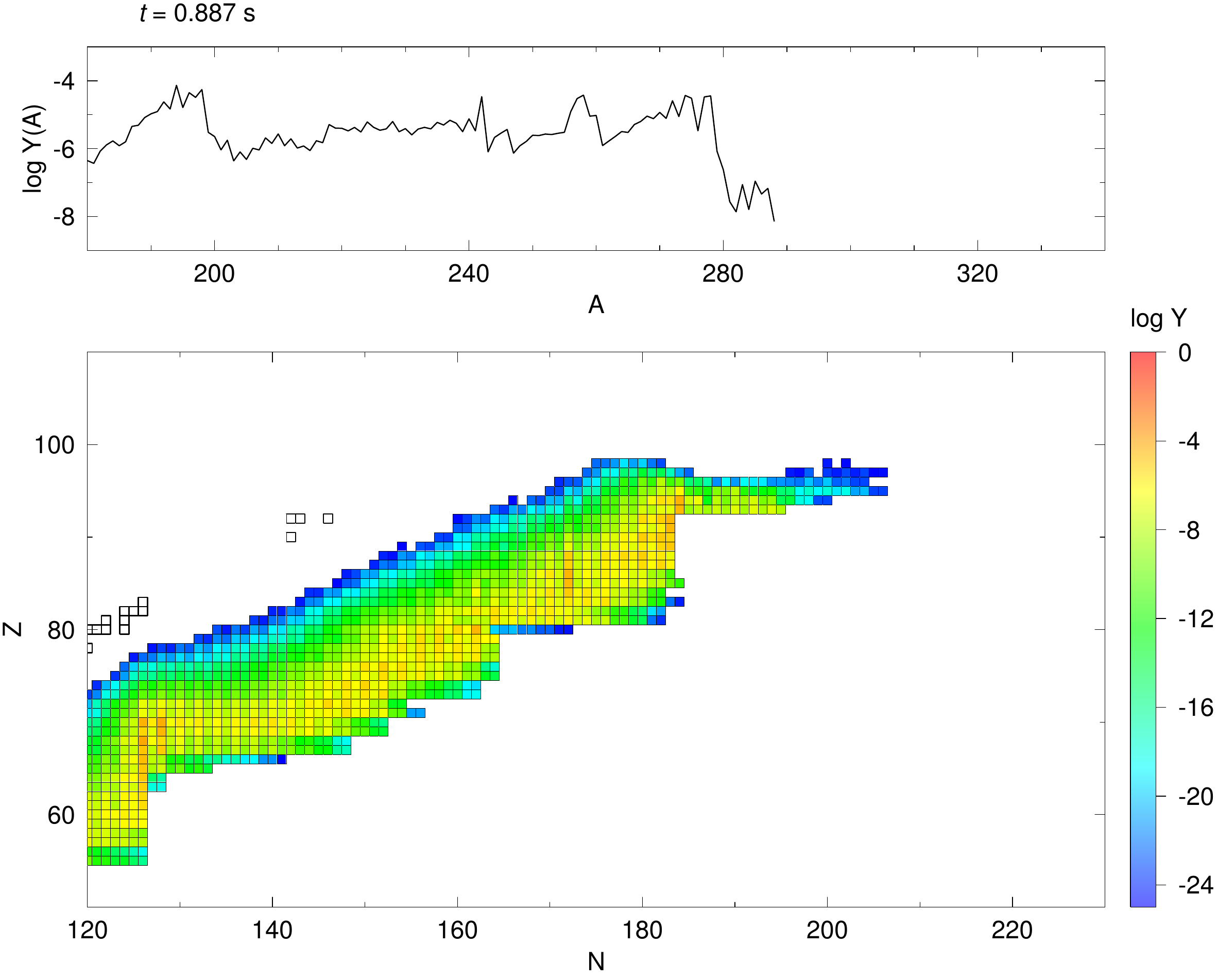}
  \caption{Results from calculations for hot (top) and cold (bottom)
    r-process conditions, utilizing fission barriers and mass
    predictions from the models ETF and FRDM. In both cases the
    abundances are shown as a function of A as well as in terms of a
    contour plot, where the abundances of nuclei are indicated by
    their color. The abundances are given at the point of neutron
    freeze-out, i.e.  when the ratio of neutrons to heavy nuclei has
    dropped down to 1. That means that on average each nucleus will
    not capture more than one neutron after this point in time.}
\label{fig:4}       
\end{figure*}

\begin{figure*}
  \centering
  \includegraphics[width=0.8\linewidth]{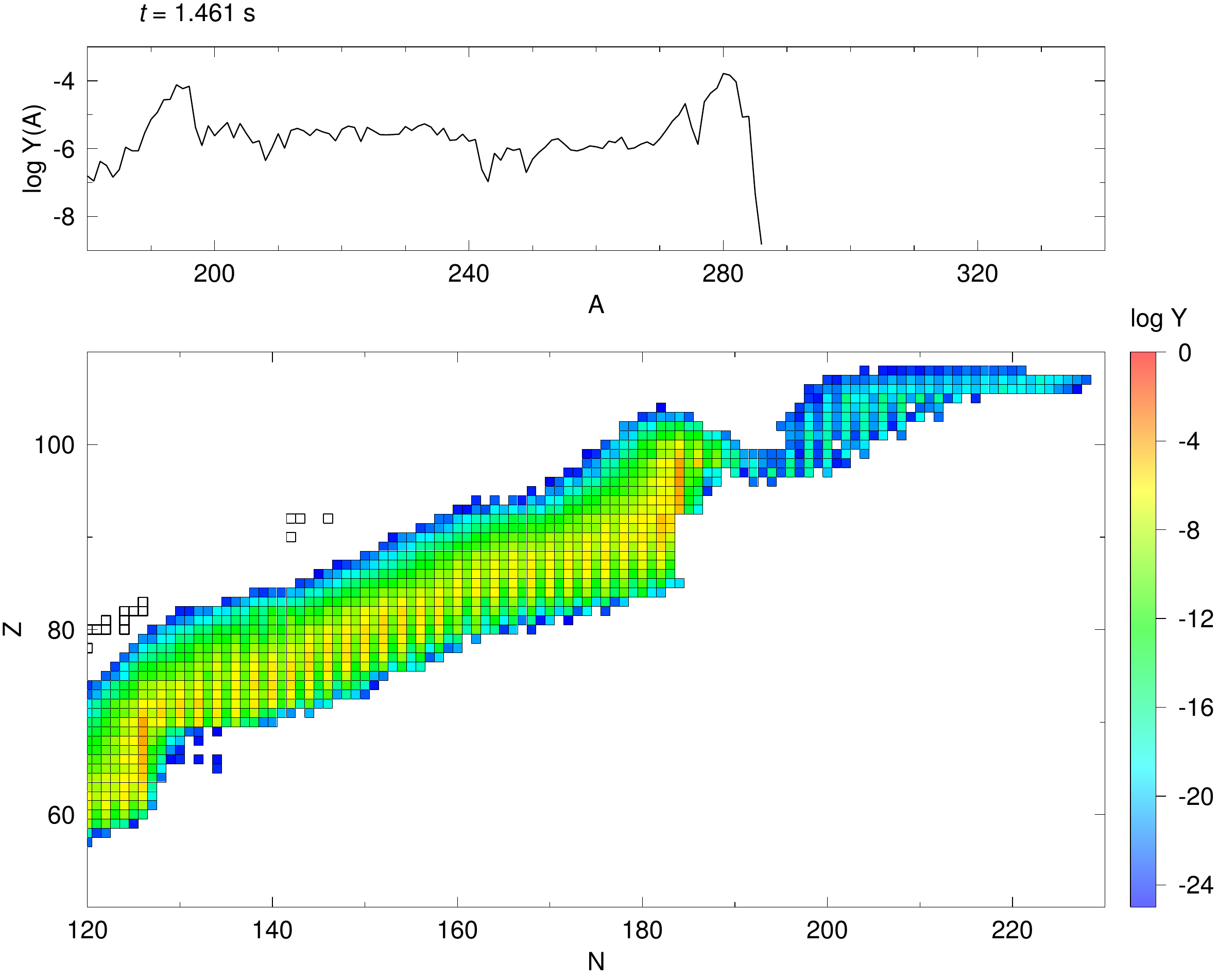}
  \includegraphics[width=0.8\linewidth]{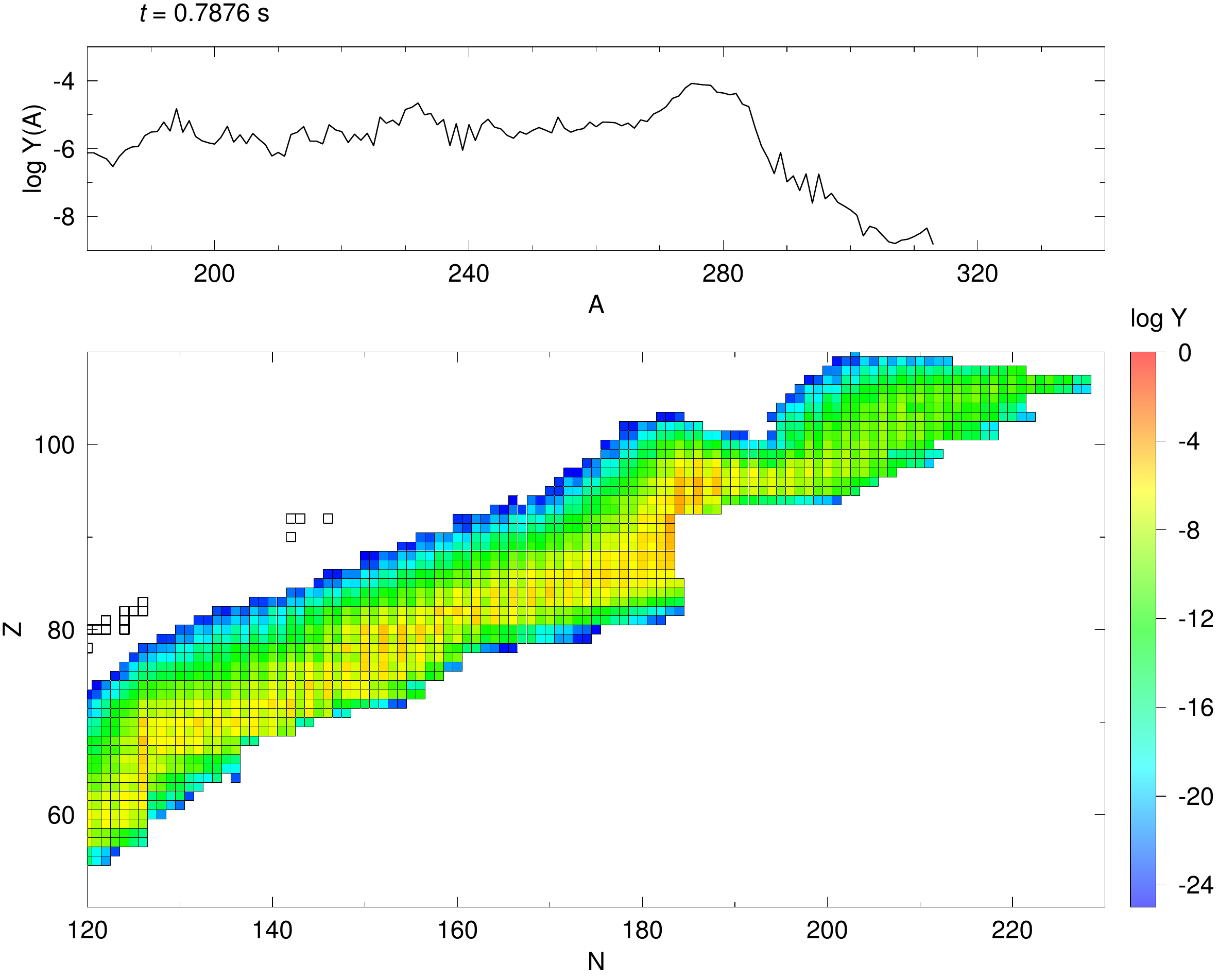}
  \caption{Same as Fig.~\ref{fig:4}, but utilizing the ETFS/ETFSI
    combination of fission barrier and mass predictions. It can be
    seen that the evolution procedes to higher masses than in
    Fig.~\ref{fig:4} and fission is less important during the working
    of the r-process. See text for a discussion.\label{fig:5}}       
\end{figure*}

Besides this more general discussion, addressing the working of the
r-process, we want to concentrate now on the focus of the present
investigation, the effect of fission and the path to superheavy
nuclei. For this purpose we have performed calculations in the hot and
cold r-process scenarios using the TF/FRDM and ETFSI/ETFSI input
sets. We focus in the following on the mass regime $A>180$ that is
relevant for our discussion. The abundance distribution at the moment
where the neutron-to-seed ratio reaches 1, marking the beginning of
neutron-capture freeze-out, is shown in figs.~{\ref{fig:4}} and
\ref{fig:5}. In each case a comparison is given between a hot
and a cold r-process, shown in Fig.~{\ref{fig:4}} for a nuclear input
set TF/FRDM for fission barriers and masses and in Fig.~{\ref{fig:5}}
for the set ETFSI/ETFSI.

We see that in both cases (hot and cold r-process) and for both sets
of nuclear input the abundances
decline substantially for nuclei $A>280$, extending in the cold case
for a few more mass units. As discussed
by~\cite{SchrammFizet:1973,Blake.Schramm:1976} in a hot r-process 
$(n,\gamma)\rightleftarrows(\gamma,n)$ equilibrium ensures a
substantial leakage into the fission channel, hindering the production
of heavier elements once
an isotopic chain is reached where neutron-induced fission can occur.
This occurs in the region $Z=95$--100 close to or
slightly beyond $N=184$, where as seen in Fig.~\ref{fig:bfsn} $B_f-S_n
< 2$~MeV. Once neutrons are exhausted, Fig.~\ref{fig:dominant} (top)
shows that spontaneous fission can become the dominating fission
channel. In the cold r-process scenario the evolution is dominated by
individual nuclei with low fission rates, allowing the production of
heavier nuclei than in the hot scenario.

\begin{figure}
  \includegraphics[width=\linewidth]{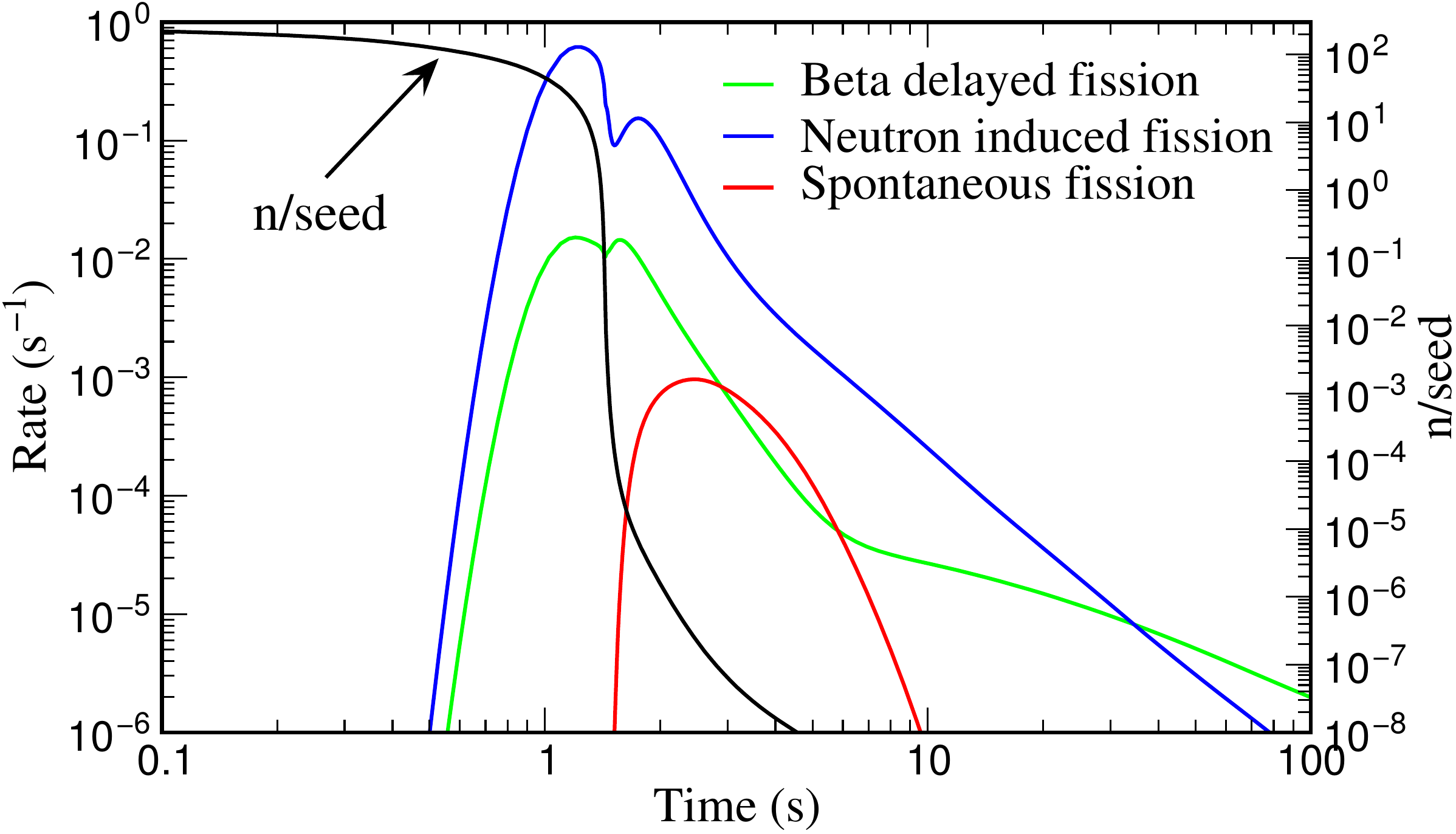}
  \includegraphics[width=\linewidth]{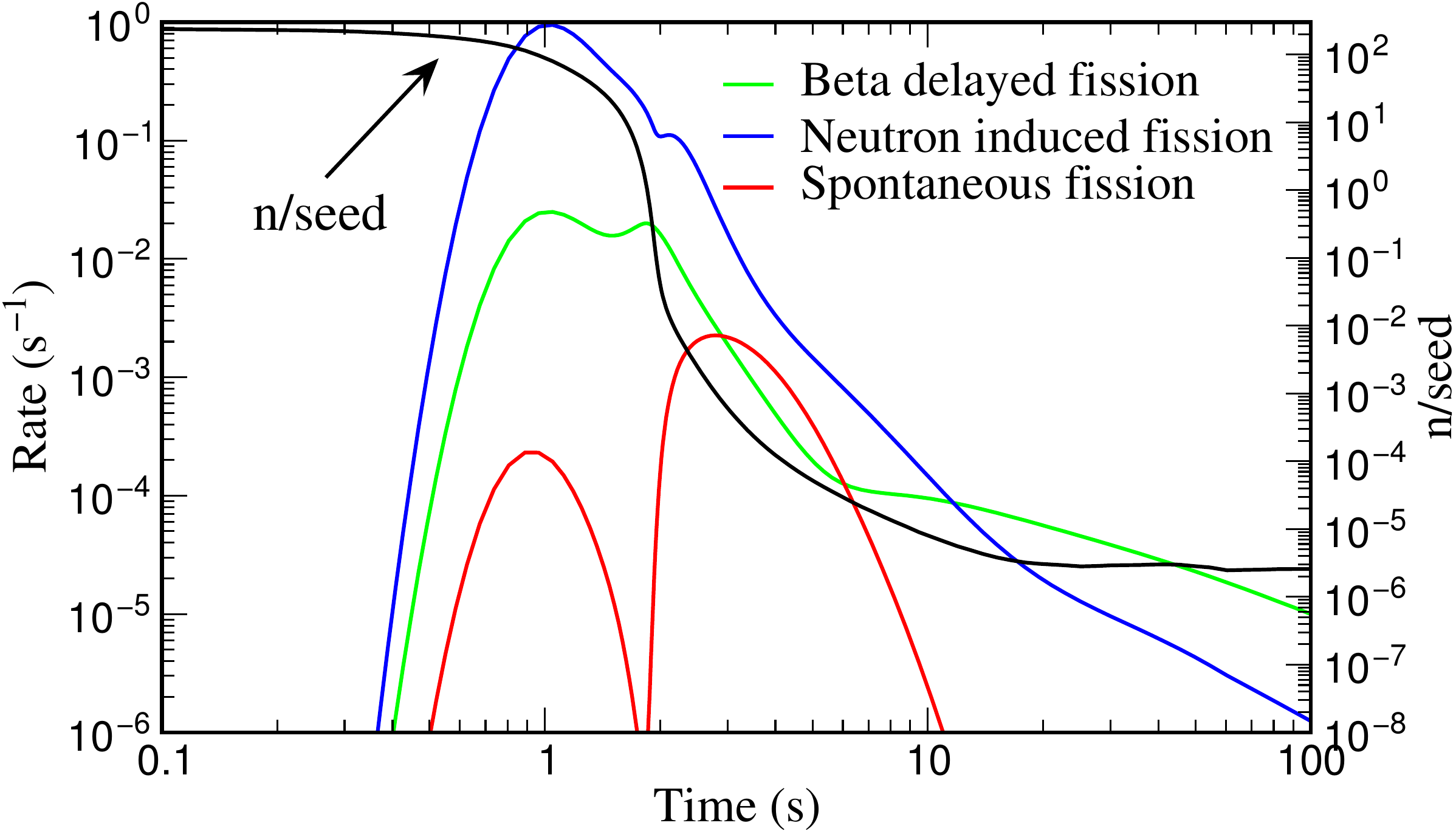}
  \caption{The evolution of fission rates for the different channels
    shown for a hot (top) and cold (bottom) r-process with the fission
    barrier/mass model selection TF/FRDM. \label{fig:4a}} 
\end{figure}

\begin{figure}
  \includegraphics[width=\linewidth]{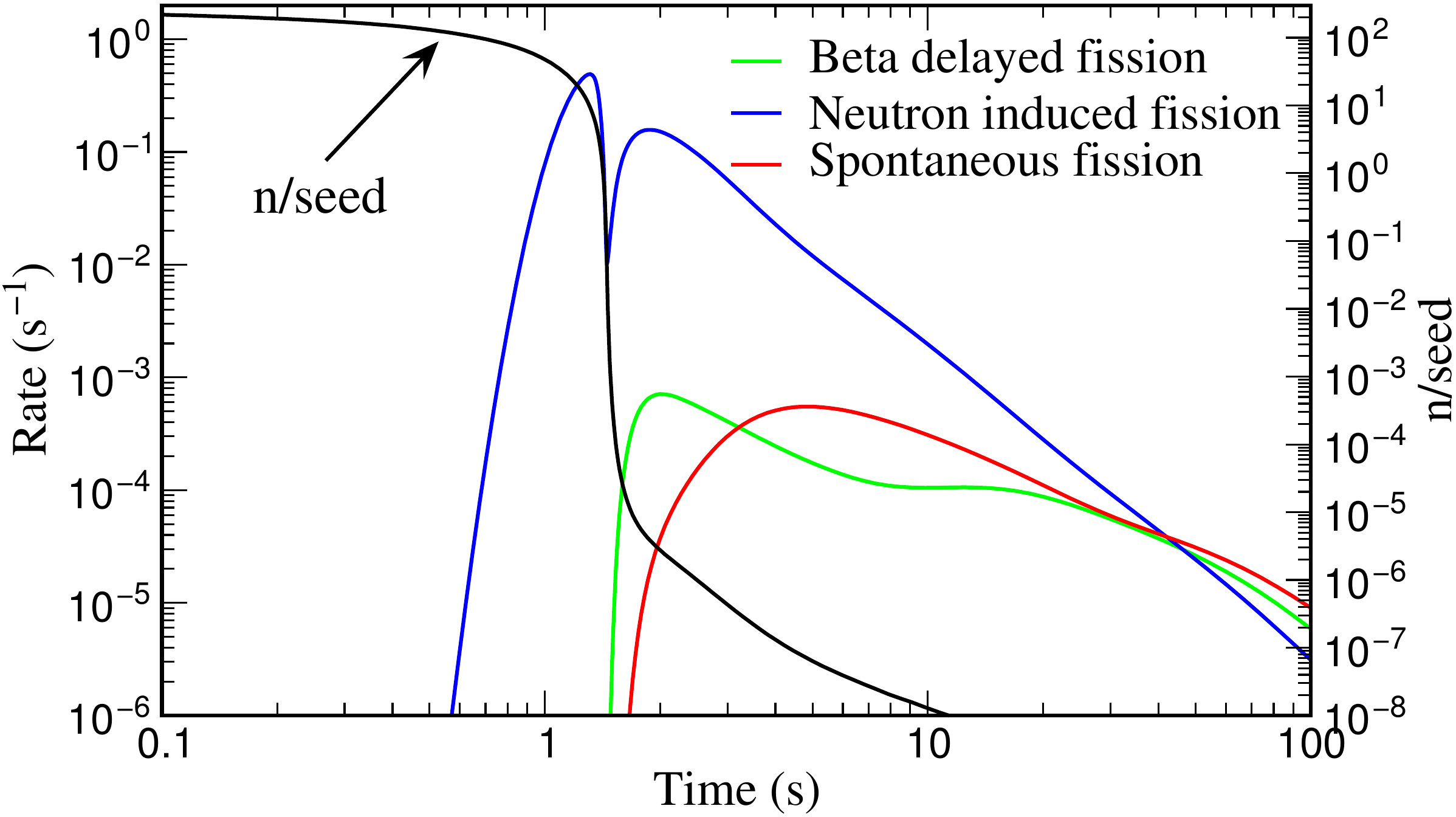}\\
  \includegraphics[width=\linewidth]{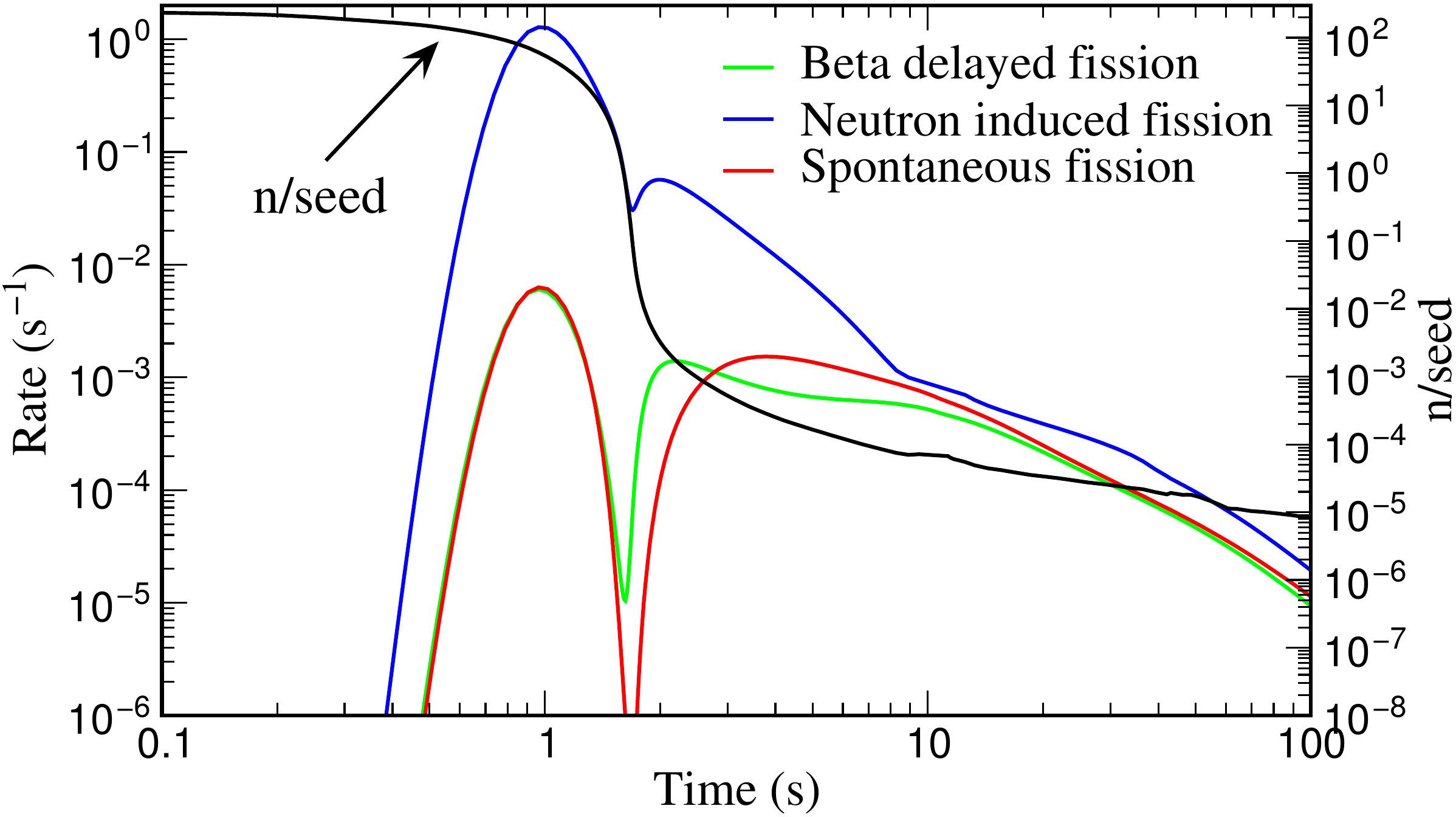}
\caption{Similar to Fig.~{\ref{fig:4a}}, the evolution of fission rates for the 
different channels shown for a hot (top) and cold (bottom) r-process, but with 
the fission barrier/mass model selection ETFSI/ETFSI.\label{fig:5a}}
\end{figure}

To explore the dependence on the nuclear input, figs.~\ref{fig:4a} and
\ref{fig:5a} show the evolution of fission rates for the different
channels for a hot and cold r-process using the fission rates and
masses TF/FRDM and ETFSI/ETFSI. There are several general features
that are worth to be mentioned. During the r-process phase, defined by a
neutron-to-seed ratio larger than one, neutron-induced fission
dominates. Once the neutron-to-seed becomes smaller than one and the
r-process freeze-out starts, the rate of neutron-induced fission
suddenly decreases. At this moment, the r-process material starts to
beta-decay to stability, producing beta-delayed neutrons via beta-delayed
neutron emission that can
induce neutron-captures but also neutron-induced fission. The latter is
particularly important for the matter accumulated at $N=184$ ($A\sim
280$) that during the beta-decay to the stability feeds the region
with $Z\sim 95$, $N\sim 175$ where neutron-induced fission can operate
again, see fig.~\ref{fig:bfsn}. This revival of neutron-induced
fission is seen in all the calculations by a second hump that is
delayed by the time scale of the successive beta decays. Once
neutron-induced is revived it sustains itself by a mechanism similar to
a chain reaction and continues to be the dominating channel till the
neutron density becomes too low at times around several tens of
seconds when beta-delayed fission and spontaneous fission
dominate. Spontaneous fission, which is dominant for nuclei closer to stability, is always less important
during the working of the r-process, but becomes larger in the
freeze-out phase, when the path moves closer to stability. In the cold
case the path is more spread out in abundances, as discussed above, and
encounters, therefore, the spontaneous fission region (closer to
stability) earlier, already before the decay back to stability sets
in.
For nuclei with the highest mass numbers (see also Fig.(\ref{fig:dominant})
beta-delayed fission can also be important at late times. 

While for TF/FRDM nuclear input we have seen that neutron-induced
fission prevents the build-up of (super)heavy elements, the situation
looks different when applying ETFSI/ ETFSI for fission barriers and
masses (see Fig.~\ref{fig:5}). This is understandable from
Fig.~{\ref{fig:bfsn}}, where we see that the region with $B_f-S_n<0$
is further removed from the drip-line (and r-process path) than for
the TF/FRDM case.  As beta-delayed fission is always less prominent
than neutron-induced fission during the r-process build-up, we also
expect matter to continue to proceed to elements heavier than those
included in the present calculation (see sect.~\ref{sec:1}).
The relative role of the three fission channels shown in
Fig.~\ref{fig:5a} is similar to that shown in Fig.~{\ref{fig:4a}}, but
with the exception that neutron-induced and beta-delayed fission do
not fully hinder the production of nuclei heavier than those included
in the present network, limited to $Z\leq110$. Another difference is
that spontaneous fission becomes more prominent in the late freeze-out
phase when the path moves closer to stability. This can be understood
as neutron-induced and beta-delayed fission do not fully prevent the
production of heavy nuclei.

It was already realized from Fig.~\ref{fig:dominant} that the area
where spontaneous fission dominates is split for the ETFSI/ETFSI case,
leaving a ``free passage'' in the vicinity of $N=184$
\cite{Panov:etal:2009}. 
This permits
beta-decay in the direction of the valley of stability. The decay path
will encounter, however, the lower of the two spontaneous fission
islands, at a charge number $Z>110$, i.e. outside the limits of the
presently used nuclear network. 

\begin{figure*}
\includegraphics[width=0.5\linewidth]{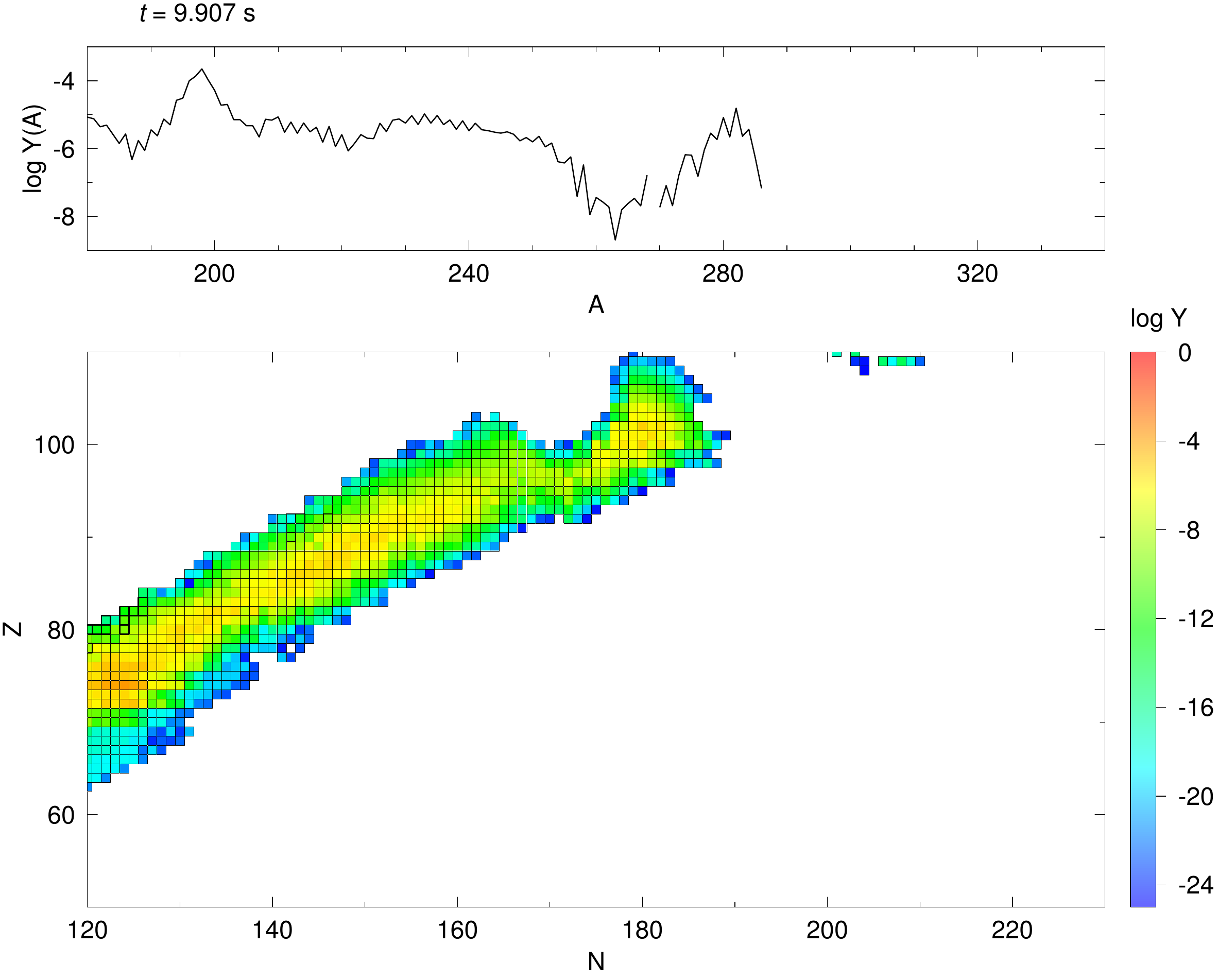}%
\includegraphics[width=0.5\linewidth]{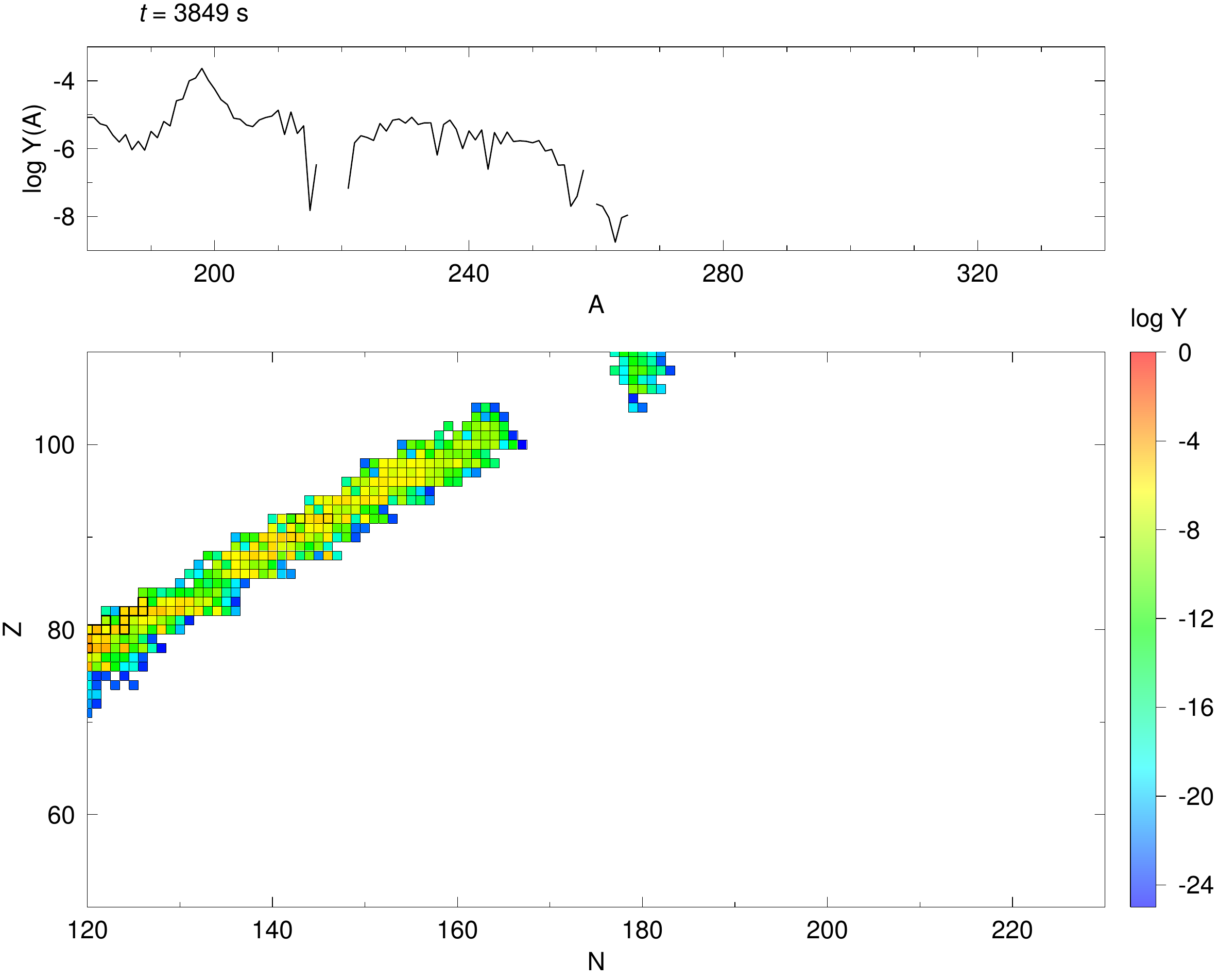}\\
\includegraphics[width=0.5\linewidth]{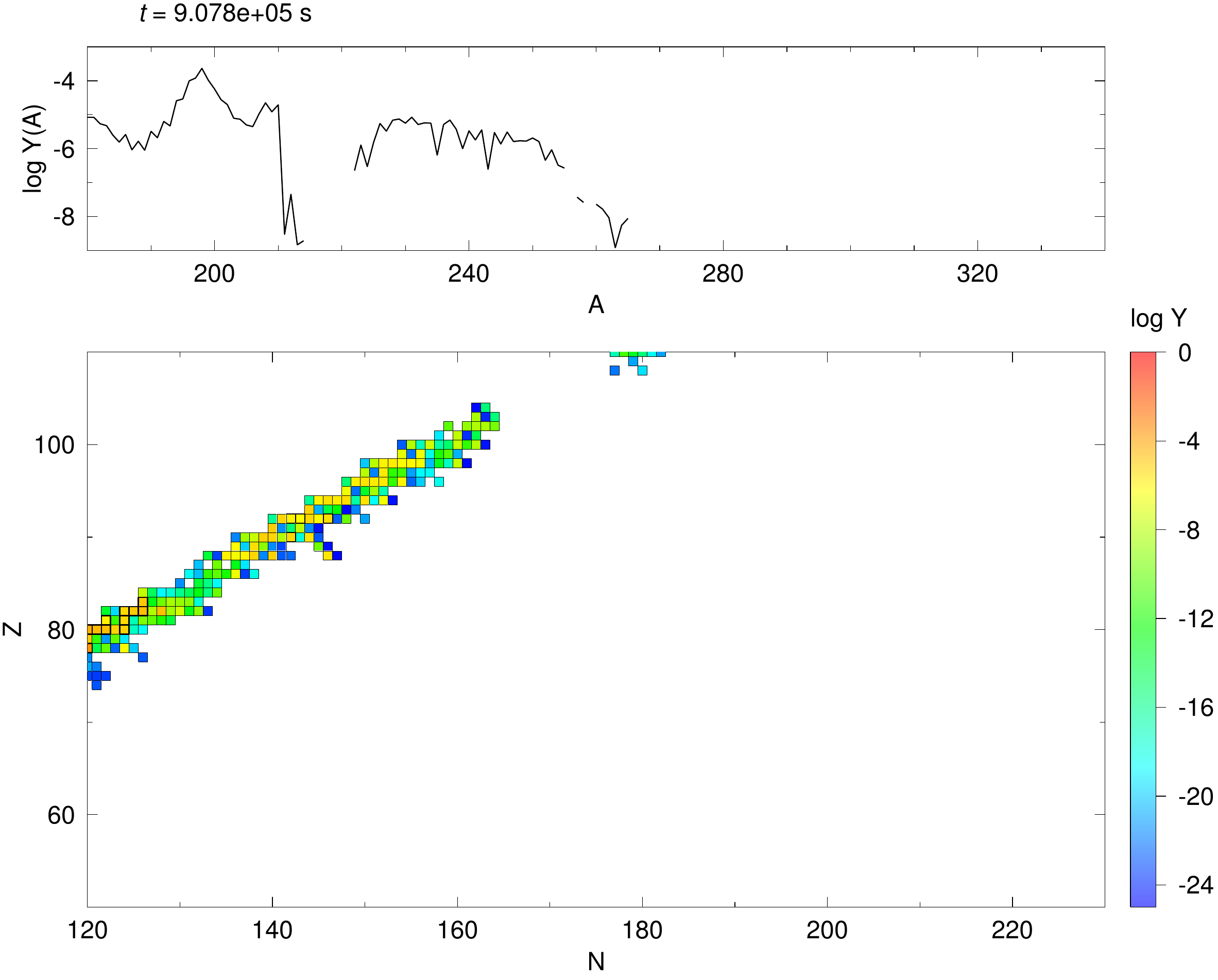}%
\includegraphics[width=0.5\linewidth]{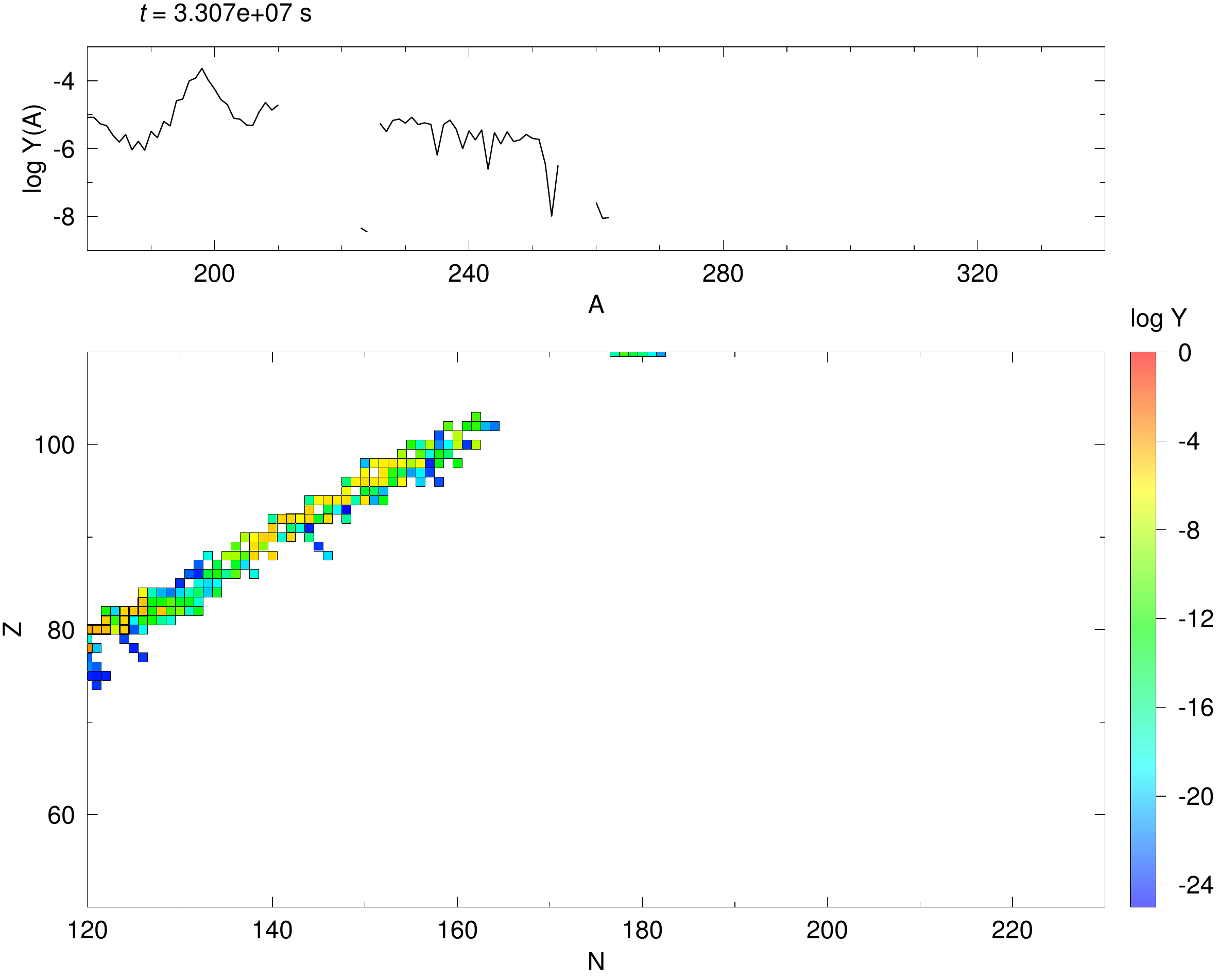}
\caption{Top and bottom figures show the abundance patterns of hot and cold 
r-process calculations with ETFST/ETFSI fission barriers and masses. The
plots refer to the latest point in time when still Z=109 nuclei are present,
or just have decayed to Z=100. These are the results of decay via the "free passage"
region of spontaneous fission shown in Fig.~{\ref{fig:dominant}}. }
\label{fig:6}       
\end{figure*}
Fig.~\ref{fig:6} shows (for the ETFSI/ETFSI combination of fission
barriers and masses) results of hot and cold r-process calculations at
that point in time when Z=109 nuclei are still present or just decayed
to Z=110, i.e.  when reaching the limits of the nuclear network
in the late decay phase.  In both calculations this happens after about
12h. Thus, when utilizing ETFSI/ETFSI fission barriers and masses,
such (superheavy) nuclei exist for this amount of time.  From
Fig.~\ref{fig:dominant} we know, however, that they will finally also
reach the lower island of spontaneous fission and not reach the valley
of stability as final destination.

As noticed above in the ETFSI/ETFSI case the r-process involves nuclei
heavier than those included in our network and consequently we cannot
determine the answer to question (iii) of the introduction, i.e. the
possibility of producing nuclei heavy enough to circumvent the region
of fission dominance also during the decay to stability. This option
is not available for the TF/FRDM case as sufficiently heavy nuclei are
never produced during the r-process.

\section{Discussion and Summary}

In this paper we examined whether superheavy elements can be
synthesized in nature by the r-process. As the r-process conditions in
astrophysical explosions are not fully understood, we utilized two
sets of calculations, relating to the so-called hot and cold environments,
where either photo-disintegrations play or do not play an important
role. For conditions with sufficient neutron-to-seed ratios to produce
actinide and heavier nuclei, we came to essentially two types of
conclusions.

For one set of fission barriers and mass predictions (TF/FRDM) the
r-process is terminated by neuton-induced and beta-delayed fission
before superheavy nuclei can be reached. For the other set
(ETFSI/ETFSI) neutron-induced and beta-delayed fission do not prevent
the build-up of superheavy nuclei even extending beyond
$Z>110$. Unfortunately, this is the edge of our network calculations,
related to the current limits of available global nuclear input.
Independent of this shortcoming, our calculations predict that nuclei
with masses of the order $A=290$ are produced and survive for about 12
hours after an r-process event (which only takes
seconds). After several beta decays also these nuclei will encounter
regions of spontaneous fission an decay to medium mass nuclei before
reaching the valley of stability.

The present investigation is another example showing the importance of
nuclear input for r-process nucleosynthesis. A reduction of the
underlying uncertainties by either experiment or theory is
desirable. This should also include an extension of the calculations
to nuclei with higher charge, $Z>110$, and neutron number to determine
if the r-process can produce progenitors with such high mass numbers
that during the decay to the stability it can circumvent the region of
fission dominance predicted by current models ($N\sim 184, Z>100$).
A first step towards this goal has been achieved in a recent
study~\cite{Erler.Langanke.ea:2012}, which shows that within the
uncertainties among different Skyrme functionals it is likely that
spontaneous-fission will hinder the production of elements beyond
$Z=120$.

\begin{acknowledgement}
We acknowledge enlightening discussions with P. Armbruster,
H. Feldmeier 
and A. Kelic-Heil. 
G.M.P. is partly supported by the Deutsche
Forschungsgemeinschaft through contract SFB 634 and the Helmholtz
International Center for FAIR within the framework of the LOEWE
program launched by the state of Hesse. 
I.V.P. is supported by the Russian goverment grant 11.G34.31.0047.
P.-G.R. is supported by the BMBF under contract 06 ER 9063. 
F.-K.T. research is funded by the Swiss National Science Foundation
(SNF) and a Humboldt Research Award.
This joint collaboration benefitted from the EU FP7 infrastructure grant
ENSAR/THEXO, the collaborative research program Eurogenesis (ESF), 
the Helmholtz Association through the Nuclear Astrophysics Virtual Institute 
(VH-VI-417), and the SNF program SCOPES for joint research with Eastern
Europe.
\end{acknowledgement}

%
%

\end{document}